\documentclass[prb,aps,twocolumn,superscriptaddress,showpacs]{revtex4-1}

\usepackage{graphicx}
\usepackage{epstopdf}

\begin{document}

\title{Spin-orbit transitions in $\alpha$ and $\gamma$-CoV$_{2}$O$_{6}$}

\author{F. Wallington}
\affiliation{School of Physics and Astronomy, University of Edinburgh, Edinburgh EH9 3FD, UK}
\author{A. M. Arevalo-Lopez}
\affiliation{Centre for Science at Extreme Conditions, University of Edinburgh, Edinburgh EH9 3FD, UK}
\affiliation{School of Chemistry, University of Edinburgh, Edinburgh EH9 3FD, UK }
\author{J. W. Taylor}
\affiliation{ISIS Facility, Rutherford Appleton Laboratory, Chilton, Didcot OX11 0QX, UK}
\author{J. R. Stewart}
\affiliation{ISIS Facility, Rutherford Appleton Laboratory, Chilton, Didcot OX11 0QX, UK}
\author{V. Garcia-Sakai}
\affiliation{ISIS Facility, Rutherford Appleton Laboratory, Chilton, Didcot OX11 0QX, UK}
\author{J. P. Attfield}
\affiliation{Centre for Science at Extreme Conditions, University of Edinburgh, Edinburgh EH9 3FD, UK}
\affiliation{School of Chemistry, University of Edinburgh, Edinburgh EH9 3FD, UK }
\author{C. Stock}
\affiliation{School of Physics and Astronomy, University of Edinburgh, Edinburgh EH9 3FD, UK}
\affiliation{Centre for Science at Extreme Conditions, University of Edinburgh, Edinburgh EH9 3FD, UK}

\date{\today}

\begin{abstract}

$\gamma$-triclinic and $\alpha$-monoclinic polymorphs of CoV$_{2}$O$_{6}$ are two of the few known transition metal ion based materials that display stepped $1/3$ magnetization plateaus at low temperatures.  Neutron diffraction [M. Markkula \textit{et al.} Phys. Rev. B {\bf{86}}, 134401 (2012)], x-ray dichroism [N. Hollmann \textit{et al.} Phys. Rev. B {\bf{89}}, 201101(R) (2014)], and dielectric measurements [K. Singh \textit{et al.} J. Mater. Chem. {\bf{22}}, 6436 (2012)] have shown a coupling between orbital, magnetic and structural orders in CoV$_{2}$O$_{6}$.  We apply neutron inelastic scattering to investigate this coupling by measuring the spin-orbit transitions in both $\alpha$ and $\gamma$ polymorphs.  We find the spin-exchange and anisotropy in monoclinic $\alpha$-CoV$_{2}$O$_{6}$ to be weak in comparison with the spin-orbit coupling $\lambda$ and estimate an upper limit of $|J/\lambda| \sim$ 0.05.  However, the spin exchange is larger in the triclinic polymorph and we suggest the excitations are predominately two dimensional.  The local compression of the octahedra surrounding the Co$^{2+}$ ion results in a direct coupling between higher energy orbital levels, the magnetic ground state, and elastic strain.  CoV$_{2}$O$_{6}$ is therefore an example where the local distortion along with the spin-orbit coupling provides a means of intertwining structural and magnetic properties.  We finish the paper by investigating the low-energy magnetic fluctuations within the ground state doublet and report a magnetic  excitation that is independent of the local crystalline electric field.  We characterize the temperature and momentum dependence of these excitations and discuss possible connections to the magnetization plateaus.

\end{abstract}

\maketitle

\section{Introduction}

In magnets based upon a triangular arrangement, conventional mean field phases are often suppressed allowing new states of matter to be studied.~\cite{Collins97:75}    Because of the local geometry, these  systems are intrinsically low dimensional (Ref. \onlinecite{Stock09:103}) and often provide a framework to study one-dimensional physics which has led to the discovery and study of spinon excitations in $S=1/2$ chains (Refs. \onlinecite{Tennant93:73,Lake05:4,Piazza14:11}) and the Haldane gap in $S=1$ magnets (Refs. \onlinecite{Haldane83:50,Buyers86:56,Kenzelmann02:66,Kenzelmann02:66,Xu96:54}).  Unconventional dynamics and phases can also result and possibilities include spin-liquid phases and nematic interactions.~\cite{Balents10;464,Nakatsuji05:309,Nambu06:75,Stock10:105,Bhattacharjee06;74}  The physics and phases are often strongly analogous to strongly correlated electronic systems, but in a context that is more amenable to theory.  Triangular magnets are also the focus of research in multiferroics due to the natural coupling between magnetic and structural orders because of the intrinsic geometry.~\cite{Eerenstein06:442,Cheong07:6}

An important discovery in condensed matter physics has been the Quantum Hall Effect where the Hall conductance displays plateaus as a function of field.  While the quantum Hall effect is an electronic phenomena in metals, an analogy has been predicted to exist in insulating spin-chains where it was suggested that the magnetization will display plateaus as a function of field.~\cite{Oshikawa97:78}  Since this prediction there have been several systems which have been found to display clear plateaus in the magnetization including insulating Cu$_{3}$(CO$_{3}$)$_{2}$(OH)$_{2}$ (Ref. \onlinecite{Kikuchi05:94}), Ca$_{3}$Co$_{2}$O$_{6}$ (Refs. \onlinecite{Kageyama97:66,Kageyama97:66_2,Hardy04:70}), Ca$_{3}$CoRhO$_{6}$ (Ref. \onlinecite{Niitaka01:87,Samo02:65,Hardy03:15}), and Sr$_{3}$HoCrO$_{6}$ (Ref. \onlinecite{Hardy06:74}).  While the analogy between the plateaus and topological phases, such as the Quantum Hall Effect, is interesting, there have been other theories for the magnetization plateaus including ``quantum tunnelling of the magnetization" (Ref. \onlinecite{Maignan04:14}), field driven transitions in the magnetic structure (Ref. \onlinecite{Kudasov06:96,Fishman11:106,Markkula12:86}), and ``dimer-monomer" model applied to Azurite (Ref. \onlinecite{Okamoto03:15,Rule08:100}).  It is therefore important to study magnetic systems displaying plateaus in an attempt to understand the broader mechanism for this unusual phenomena. 

\begin{figure}[t]
\includegraphics[width=7cm] {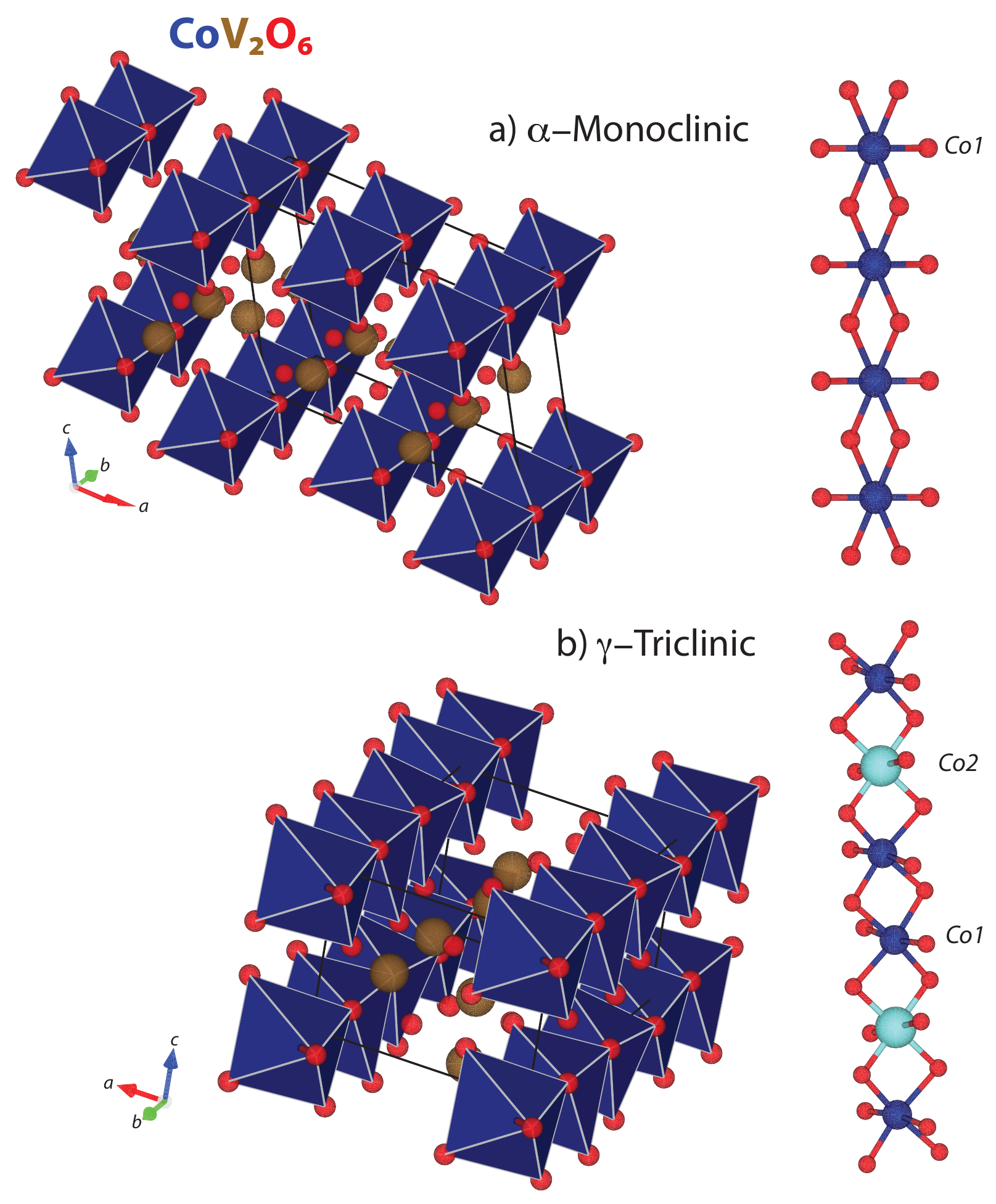}
\caption{\label{structure}  The two structural polymorphs of CoV$_{2}$O$_{6}$ brannerites based upon octahedra of oxygen and cobalt. (a) illustrates the monoclinic ($\alpha$) variant.  $(b)$ shows the triclinic polymorph.  The chain like structure is also illustrated with the two Co$^{2+}$ sites highlighted for the triclinic variant (labelled as \textit{Co1} and \textit{Co2}).}
\end{figure}

In this paper, we investigate the magnetic properties of powders of CoV$_{2}$O$_{6}$ which display $1/3$ magnetization plateaus.   CoV$_{2}$O$_{6}$ has two published structural polymorphs with one having a monoclinic unit cell ($\alpha$; space group $C2/m$, $a=9.289$ \AA, $b=3.535$ \AA, $c=6.763$ \AA, and $\beta=112.64^{\circ}$) and the other being triclinic ($\gamma$; space group $P\overline{1}$, $a=7.164$ \AA, $b=8.872$ \AA, $c=4.806$ \AA, $\alpha=90.29^{\circ}$, $\beta=93.66^{\circ}$, $\gamma=102.05^{\circ}$).  The two structures are illustrated in Fig. \ref{structure}.  Both polymorphs show $1/3$ magnetization plateaus at low temperatures in the magnetically ordered phases with the plateaus being more pronounced and extended in magnetic field in the monoclinic polymorph over the triclinic variant.~\cite{Lenertz11:115}  While these plateaus exist in the magnetically ordered phase, they are washed out and disappear quickly with temperature well below T$_{N}$ as clearly shown in the $\alpha$-monoclinic polymorph.

Because of the clear and well separated plateaus at accessible fields, monoclinic $\alpha$-CoV$_{2}$O$_{6}$ has been the focus of a number of investigations both experimentally and theoretically.  This material possesses only a single Co$^{2+}$ site which is linked in a geometry that can be referred to as a chain consisting of edge sharing octahedra (Fig. \ref{structure} $a$).   This contrasts with the case of triclinic $\gamma$-CoV$_{2}$O$_{6}$ (Fig. \ref{structure} $b$) which has two different Co$^{2+}$ sites in a 2:1 ratio with differing local bonding geometries.   Theoretical investigations of monoclinic $\alpha$-CoV$_{2}$O$_{6}$ have predicted an exchange along the chain of $\sim$ 3 meV $\sim$ 30 K and this has been used to successfully model the magnetization plateaus.~\cite{Saul13:87}  However, other studies have suggested the importance of local single-ion crystal fields acting on the Co$^{2+}$ sites in CoV$_{2}$O$_{6}$.~\cite{Hollmann14:89}  

While understanding the origin of the unusual magnetization plateaus is a central reason for studying CoV$_{2}$O$_{6}$, the polymorphs of this system also display a range of properties illustrating a coupling between structural and magnetic order.   Neutron diffraction has found a large magnetostriction with temperature (Ref. \onlinecite{Marrkkula12:192}) and also magnetic field (Ref. \onlinecite{Markkula12:86}).  Dielectric measurements (Ref. \onlinecite{Singh12:22}) have further found evidence for coupling between magnetic and dielectric constants with the application of an applied field.   There is also a strong orbital contribution to the magnetic ground state as highlighted by recent x-ray studies.~\cite{Hollmann14:89}    The goal of this study is to understand the origin of this coupling between structural and magnetic properties.

Given the Co$^{2+}$ octahedral environment allows an orbital degeneracy, there are several energy scales to consider which potentially couple structural and magnetic properties.  These include the crystalline electric field, spin-orbit coupling, and spin exchange through direct or superexchange mechanisms.~\cite{Kim12:85}  The problem is potentially complicated by the fact that the spin-orbit coupling has a relatively small energy scale for Co$^{2+}$ in an octahedral environment and potentially this is of the same order as the spin superexchange.~\cite{Cowley13:88,Cowley73:6,Kant08:78}  To understand the relative energy scales of these contributions to the magnetic Hamiltonian, we discuss neutron inelastic scattering results of powders from both polymorphs $(\alpha,\gamma)$-CoV$_{2}$O$_{6}$ over a broad dynamic range in energy.  These measurements reveal low-energy spin-orbit excitations sensitive to the crystalline electric field imposed by the structure.  The exchange constants between Co$^{2+}$ are small in comparison to this crystalline electric field terms in the Hamiltonian.  CoV$_{2}$O$_{6}$ therefore represents a case where magnetism and structure are coupled through single-ion and local crystalline electric field effects.  We also observe a very low-energy excitation which decays rapidly with temperature and we suggest that these excitations are related to the plateaus.   This study therefore defines the energy scales associated with the magnetism in $(\alpha,\gamma)$-CoV$_{2}$O$_{6}$.

The paper is divided into five sections including this introduction (section $I$).  We first discuss the experiments and the sample preparation (section $II$) followed by an outline of the single ion crystal field theory of Co$^{2+}$ in an octahedral crystalline electric field (section $III$).  We then discuss the magnetic and orbital excitations in monoclinic $\alpha$-CoV$_{2}$O$_{6}$ in terms of this theory and then compare the results to triclinic $\gamma$-CoV$_{2}$O$_{6}$ (section $IV$) and finish with a summary and conclusions (section $V$).

\section{Experimental Details}

Powder CoV$_2$O$_6$ was prepared by a solid state reaction from vanadium oxide (V$_2$O$_5$) and cobalt acetate tetrahydrate (C$_4$H$_6$CoO$_4\cdot4$H$_2$O). Stoichiometric quantities were ground in an agate mortar before being pelletised and heated to 650$^{\circ}$C for 16 hours. The powder was reground and pelletised for a second heating of 725$^{\circ}$C for 48 hours. This was followed by quenching in liquid nitrogen to form the monoclinic polymorph and slow cooling to form the triclinic phase.  For each sample, the structure was confirmed with x-ray diffraction.  Further details regarding the sample preparation can be found in Ref. \onlinecite{Marrkkula12:192}.  Sample masses for the triclinic and monoclinic polymorphs were 21.8 g and 34.8 g, respectively.

Inelastic neutron scattering was performed using the MARI and IRIS spectrometers at ISIS (Didcot, UK).  On MARI, a Gd fermi chopper was used to fix the incident energy and thick disc and nimonic choppers were used to suppress the high-energy background.  Incident energies of 10, 60, and 150 meV were used with Fermi chopper frequencies of 250, 350, and 350 Hz respectively.  The disk and nimonic choppers were always run at 50 Hz, synchronised with the main proton pulse on the target.  The energy resolution at the elastic line was measured with a vanadium standard to be 0.3, 2.5, and 8.1 meV for the 10, 60, and 150 meV configurations respectively.

To obtain higher resolution measurements of the low (less than 2 meV) energy excitations, experiments were performed on the indirect spectrometer IRIS.   The final energy was fixed, using cooled pyrolytic graphite (002) analyzers, to be E$_{f}$=1.845 meV to obtain an energy resolution of 17.5 $\mu eV$ at the elastic line.  In all experiments, the sample was cooled in a closed cycle refrigerator.

\section{Neutron scattering and Cobalt in a crystal field}

Before presenting the experimental results, we first review the single ion theory of Co$^{2+}$ in an oxygen octahedra and then apply this to powder averaged neutron spectra from both the monoclinic and triclinic polymorphs of CoV$_{2}$O$_{6}$.    The cobalt ion, Co$^{2+}$ has a $3d^{7}$ electronic configuration.  The free ion (in the absence of any crystalline electric field) has been outlined in several review works where it has been shown that the ground state corresponds to $^{4}F$ (i.e. $S=3/2$ and $L=2$) with the first excited state being $^{4}P$ separated in energy by a large energy scale of several $eV$.~\cite{Khomskii:book,Abragam:book,Ballhausen:book,McClure59:9}  To understand the excitations spectrum we consider this $^{4}F$ level as the ground state and that the $^{4}P$ states to be inaccessible on the energy range of the experiment and the temperature scale of interest.  

In this section, we consider 4 terms in the Hamiltonian acting on this $^{4}F$ states defined as $H_{tot}=H_{0} + (H_{1}+H_{2}+H_{3})$ where $H_{0}$ is the dominant cubic crystalline electric field, and the $H_{1,2,3}$ are the weaker axial and in-plane distortions of the octahedra ($H_{1,2}$) and then the molecular field caused by magnetic order $H_{3}$.  These later terms are treated as a perturbation to the large cubic crystalline electric field $H_{0}$.  

\subsection{Model Single-Ion Hamiltonian}

\begin{figure}[t]
\includegraphics[width=9cm] {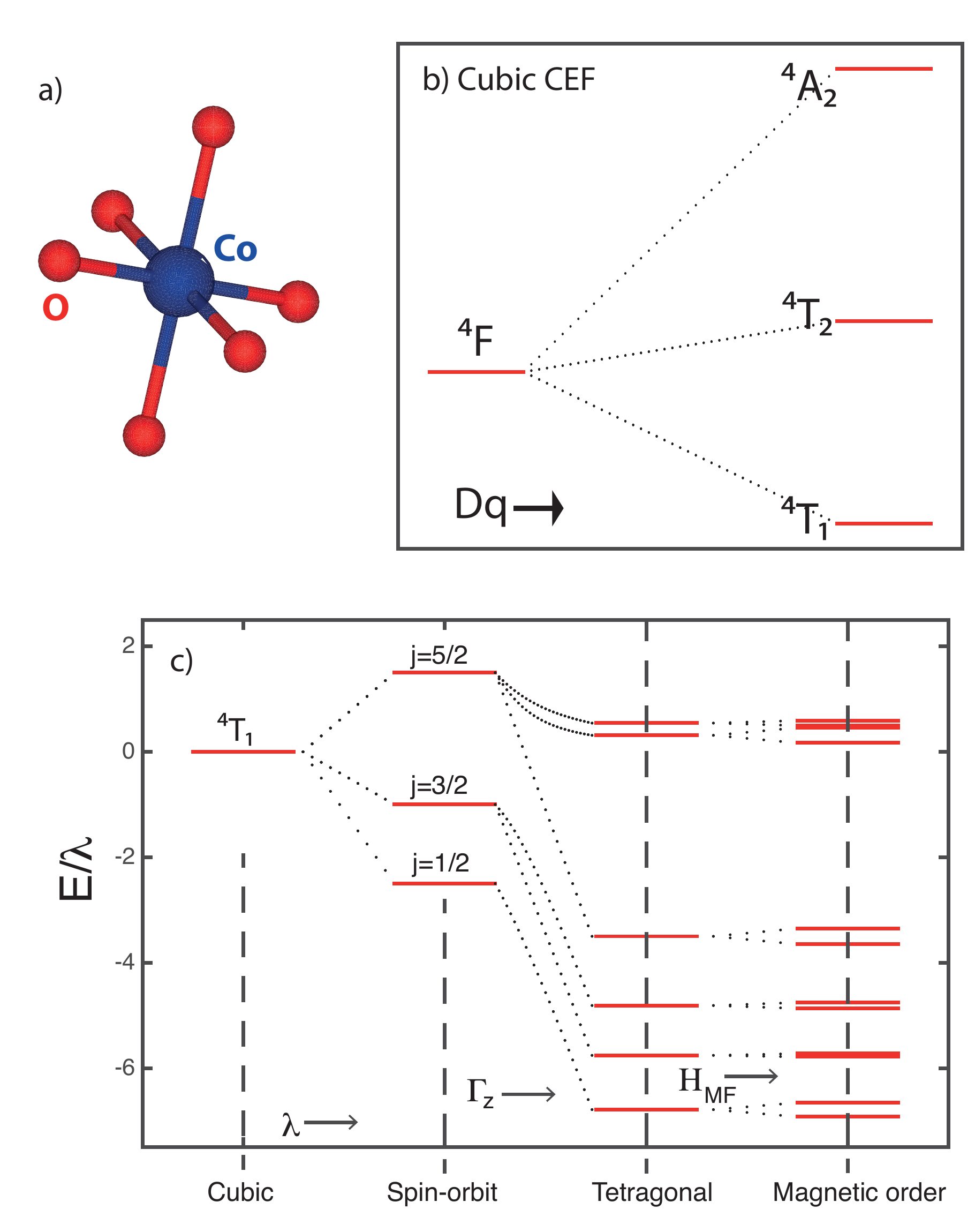}
\caption{\label{cef} (a) illustrates a plot of the Co$^{2+}$ ion surrounded by oxygen ions in CoV$_{2}$O$_{6}$ used here and how this crystal  splits the $^{4}F$ free ion state of Co$^{2+}$.  (b) The calculated energy variation as the parameters $Dq$ (cubic crystal field), $\lambda$ (spin-orbit), $\Gamma_{z}$ (tetragonal distortion), and $H_{MF}$ (magnetic molecular field) are varied.}
\end{figure}

\subsubsection{$d$ orbitals in a cubic crystal field and new basis}

The starting point for understanding the single-ion physics is to consider a local undistorted octahedra of oxygen atoms surrounding a Co$^{2+}$ ion.   This is the high temperature structure of rocksalt CoO and we use this as a benchmark to estimate the important energy scales.  In the absence of mixing between the $^{4}F$ ground state and the excited levels, (such as $^{4}P$) the Hamiltonian can be written in terms of Steven's operators ($O_{4}^{0}$ and $O_{4}^{4}$ defined in Ref. \onlinecite{Hutchings65:16}) as,

\begin{eqnarray}
H_{0}=B_{4} (O_{4}^{0}+5O_{4}^{4}).
\label{cef_cub}
\end{eqnarray}

\noindent This splits the seven orbitals states of the $^{4}F$ state into two orbitals triplets ($^{4}T_{1}$ and $^{4}T_{2}$) and one singlet ($^{4}A_{2}$).  For small values of $B_{4}$, the energy splitting between the levels is $\Delta(^{4}T_{1}\rightarrow ^{4}T_{2})=460B_{4}\equiv 8Dq$ and  $\Delta(^{4}T_{2} \rightarrow\ ^{4}A_{2})=600B_{4}\equiv 10Dq$, with the $^{4}T_{1}$ being the orbital triplet ground state.~\cite{Lines63:131}  We note that we have chosen a weak crystalline electric field approach to this problem given its success in describing the orbital excitations in CoO (Ref. \onlinecite{Cowley13:88}) as well as NiO (Ref. \onlinecite{Kim11:84}).  For CoO, $10Dq$ was estimated to be $\sim$ 1 eV with a similar energy scale found in NiO.   The first orbital transition in CoO $\Delta(^{4}T_{1}\rightarrow ^{4}T_{2})$ has been measured by several techniques with neutron scattering showing a transition of $\sim$ 0.9 eV.~\cite{Larson07:99,Haverkort07:99,Kant08:78,Cowley13:88,Abragam:book}  Recent x-ray studies have found this value to be smaller in the case of $(\alpha,\gamma)$-CoV$_{2}$O$_{6}$ with $10Dq\sim 0.5$ eV, yet large enough to further corroborate the weak crystal field analysis discussed here.~\cite{Hollmann14:89}  Given the large energy scale separating the $^{4}T_{1}$ and the $^{4}T_{2}$ levels, we only consider excitations within the $^{4}T_{1}$ degenerate levels.

In this approach, we have started with the solutions to the free Co$^{2+}$ ion and treated the cubic crystalline electric field as a perturbation.  As noted in Refs. \onlinecite{Griffiths:book,Liehr63:67,McClure59:9}, this assumption is questionable as the Coulomb interaction from the cubic crystalline field is significant.  An alternate approach is to start with the $d$ orbital states in a cubic crystalline electric field giving triply degenerate $|t_{2g}\rangle$ states separated by an energy of $10Dq$ from the higher energy $|e_{g}\rangle$ states.  Applying Hunds rules to populate these with 7 electrons either gives a high $S=3/2$ or low $S=1/2$ depending on the energy scale of $10Dq$ in comparison with the Hund's coupling.   Given spectroscopic work in CoO and CoV$_{2}$O$_{6}$ discussed above, we consider the weak limit or the high $S=3/2$ case with electronic configuration $t_{2g}^{5} e_{g}^{2}$.  For this configuration there is one hole in the $|t_{2g}\rangle$ states and given there are three possible orbitals for this hole with equal energy, this ground state is an orbital triplet.~\cite{Abragam:book,Khomskii:book}  

Therefore, through the application of either a weak or strong crystalline electric field approach, we end with the same answer that the ground state in a cubic crystalline electric field is an orbital triplet.~\cite{Abragam:book,Khomskii:book}  We note, that applying this crystal field theory to the case of a tetrahedral environment, the orbital degeneracy does not exist, though coupling to higher order orbital triplets may introduce a more complex ground state (see discussion and references in Ref. \onlinecite{Decaroli15:71}). We now consider perturbations acting on this orbital triplet.

\subsubsection{Spin-orbit coupling}

A much smaller energy scale over the cubic crystalline electric field is the spin-orbit coupling written as,

\begin{eqnarray}
H_{SO}=\tilde{\lambda} \vec{L} \cdot \vec{S}=\alpha \lambda \vec{l} \cdot \vec{S}
\label{so}
\end{eqnarray}

\noindent where $\vec{L}$ and $\vec{S}$ are the orbital and spin angular momentum, respectively, and the $\lambda$ the spin-orbit coupling constant.    For this, it is convenient to consider a total angular momentum $\vec{j}=\vec{l}+\vec{s}$ with a fictitious orbital angular momentum of $l=1$ and orbital moment projection factor of $\alpha=-3/2$.~\cite{Abragam:book,Khomskii:book}  This theoretical framework differs from considerations in real-space atomic orbitals as it has been shown in Refs. \onlinecite {Kanamor_A17:177,Kanamor_B17:177} that the low-energy magnetic states are a complicated linear combination of these.   We can work back in terms of the basis states of the $^{4}F$ state as noted in Ref. \onlinecite{Lines63:131}.   


The spin-orbit coupling $\lambda$ has recently (Ref. \onlinecite{Cowley13:88}) been extracted from a dilute sample of MgO-3\%CoO to be -16 $\pm$ 3 meV which compares well with the theoretical value of -23.4 meV.~\cite{Abragam:book}  This value is significantly less than the cubic crystalline electric field strength of $\sim$ 1 eV corroborating our approach of treating this term as a perturbation to the ground state of the cubic crystalline electric field discussed above.  The energy spectra of $H_{SO}$ is shown in Fig. \ref{cef} with the ground state being a doublet with $j_{eff}=1/2$ separated by two exited levels with $j_{eff}=3/2,5/2$ with the energy difference fixed by the Lande interval rule with  $\Delta((j_{eff}={1\over 2})\rightarrow (j_{eff}={3\over 2}))=3/2 \alpha \lambda$ and $\Delta((j_{eff}={3\over 2})\rightarrow (j_{eff}={5\over 2}))=5/2 \alpha \lambda$ .  


\subsubsection{Octahedral distortions}

The local environments around the Co$^{2+}$ in both polymorphs of CoV$_{2}$O$_{6}$ are \textit{distorted} octahedra with the Co$^{2+}$ ion position in the $\alpha$ polymorph (space group No. 12 C2/m) not having a fourfold symmetry but only having $2/m$ site symmetry.  Therefore,  other terms in the Hamiltonian need to be considered.  A distortion parallel to the axis of the octahedra (either an elongation or compression) will result in a term in the Hamiltonian of the form derived from symmetry considerations.~\cite{Walter84:45}

\begin{eqnarray}
H_{1}=\Gamma_{z} \left(l_{z}^{2}-{2\over3} \right)
\label{so}
\end{eqnarray}

\noindent The constant $\Gamma_{z}$ characterises the deviation from an ideal octahedra and therefore provides a means of coupling structural distortions to the magnetic properties.  A distortion within the plane of the octahedra will result in an additional term in the Hamiltonian,

\begin{eqnarray}
H_{2}=\Gamma_{x} \left(l_{x}^{2}-l_{y}^{2} \right).
\label{mono}
\end{eqnarray}

\noindent In other compounds which undergo distortions, and where these crystalline electric field parameters have been analyzed, $\Gamma_{z}$ and $\Gamma_{x}$ have been found to be of similar magnitude as the spin-orbit coupling.  This has been studied in detail in CoF$_{2}$.~\cite{Martel68:46,Cowley73:6,Gladney66:146}  Similar to the treatment of H$_{SO}$, we consider these terms a perturbation on the original $^{4}T_{1}$ states and the large cubic crystalline electric field term in the Hamiltonian.  As implied by Kramers theorem, all of these terms in the Hamiltonian originating from localized crystal field effects gives an energy spectrum consisting of doublets.

\subsubsection{Magnetic order and Anisotropy}

The final perturbing term we consider in the Hamiltonian is the molecular field on each site as a result of magnetic order.  Following Ref. \onlinecite{Buyers86:56}, we also include in this term effects due to anisotropy originating from dipolar effects due to the distorted octahedra and exchange anisotropy.    While it is difficult to characterize exchange anisotropy using powdered averaged data, we discuss possible scenarios later in the paper in comparison to other insulating Co$^{2+}$ materials.

\begin{eqnarray}
H_{3}=H_{MF}S^{z}
\label{so}
\end{eqnarray}

\noindent This term in the Hamiltonian breaks time reversal symmetry and therefore splits the Kramers doublets originating from the crystal field terms described above.  This is illustrated in Fig. \ref{cef} $(c)$ as the last term which splits the doublets.   Given the magnetic structure reported in Refs. \onlinecite{Markkula12:86,Lenertz11:115}, we note that in the magnetically ordered state there is a non cancellation of the molecular field on each Co$^{2+}$ site implying that a localized magnetic field is present in the low-temperature ordered phase.

\subsection{Neutron scattering intensities}

Neutron scattering is sensitive to magnetic dipole transitions with intensities for transitions from the ground state ($|0\rangle$) to an excited state ($|\alpha\rangle$) given by (in the absence of large thermal population),

\begin{eqnarray}
I(|0\rangle \rightarrow |\alpha\rangle) \sim \sum_{i=x,y,z} |\langle 0|M_{i}|\alpha\rangle|^{2},
\label{int}
\end{eqnarray}

\begin{figure}[t]
\includegraphics[width=8.7cm] {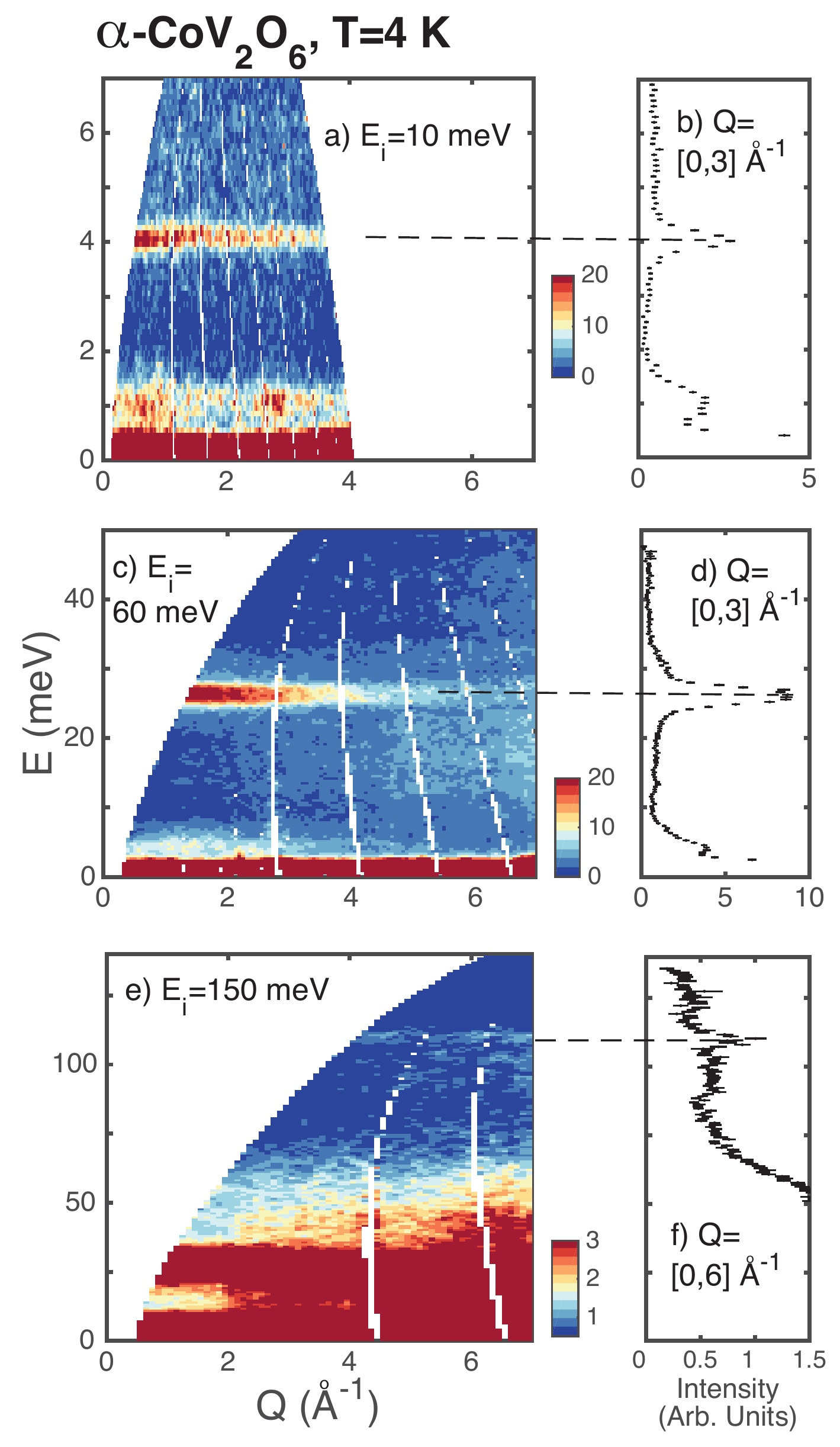}
\caption{\label{mono_summary}  A summary of the T=4 K neutron inelastic scattering results performed on $\alpha$-CoV$_{2}$O$_{6}$ (monoclinic polymorph) obtained on MARI with incident energies of E$_{i}$=$(a)$ 10, $(b)$ 60, and $(c)$ 150 meV.  All data were obtained in the magnetically ordered phase.  We note the low energy scattering at $\sim$ 1 meV is discussed at the end of the paper.}
\end{figure}

\noindent where $\vec{M}=\vec{L}+2\vec{S}$.  The ordered magnetic moment measured with neutrons is $\mu=|\langle0| M |0\rangle|\equiv |\langle L \rangle + 2 \langle S \rangle|$.  The value of the magnetic moment therefore can provide a means of characterizing the orbital contribution and the mixing of the ground state doublet with higher energy multiplets.

Eqn. \ref{int} illustrates that neutrons obey the selection rules that $\Delta m_{z}=0,\pm1$. By diagonalizing the spin-orbit/crystal field Hamiltonian above for a 12 $\times$ 12 matrix (in terms of the $|l=1, m_{l};s=3/2, m_{s}\rangle$ basis states), the expected intensities can be calculated. Based on this relation for the neutron scattering intensity, for an undistorted octahedra, the strongest excitations are those within the ground state $j_{eff}=1/2$ doublet and also between the ground state $j_{eff}=1/2$ doublet and first excited state $j_{eff}=3/2$ manifold of states.  Transitions to $j_{eff}=5/2$ are not present.  However, distorting the octahedra through including additional crystal field terms (H$_{1,2}$) discussed above does allows further excited states including those of the high energy $j_{eff}=5/2$ manifold of states to be allowed as these are mixed in with the ground state.

\begin{figure}[t]
\includegraphics[width=8.7cm] {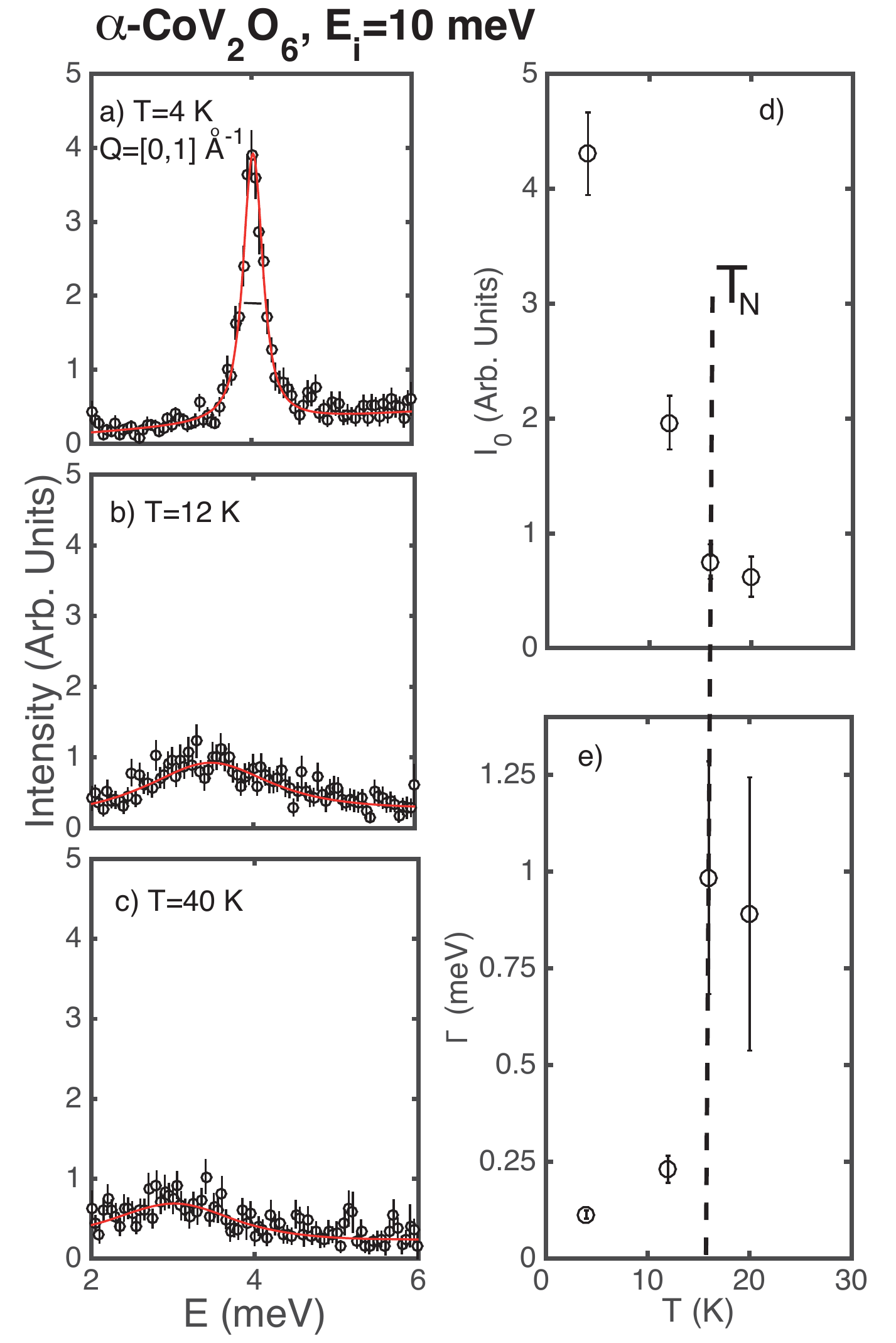}
\caption{\label{summary_10meV}  The temperature dependence of the 4 meV peak measured on MARI. $(a)$-$(c)$ show a constant momentum scan integrating over $Q=[0,1]$\AA$^{-1}$ demonstrating the disappearance of the peak above T$_{N}$.  $(d)$ and $(e)$ show how the fitted intensity and line width vary with temperature further confirming that the peak is directly died to the antiferromagneticaly ordered phase.  We assign this transition to an excitation within the lowest energy $j_{eff}={1\over 2}$ doublet.}
\end{figure}

\begin{figure}[t]
\includegraphics[width=8.7cm] {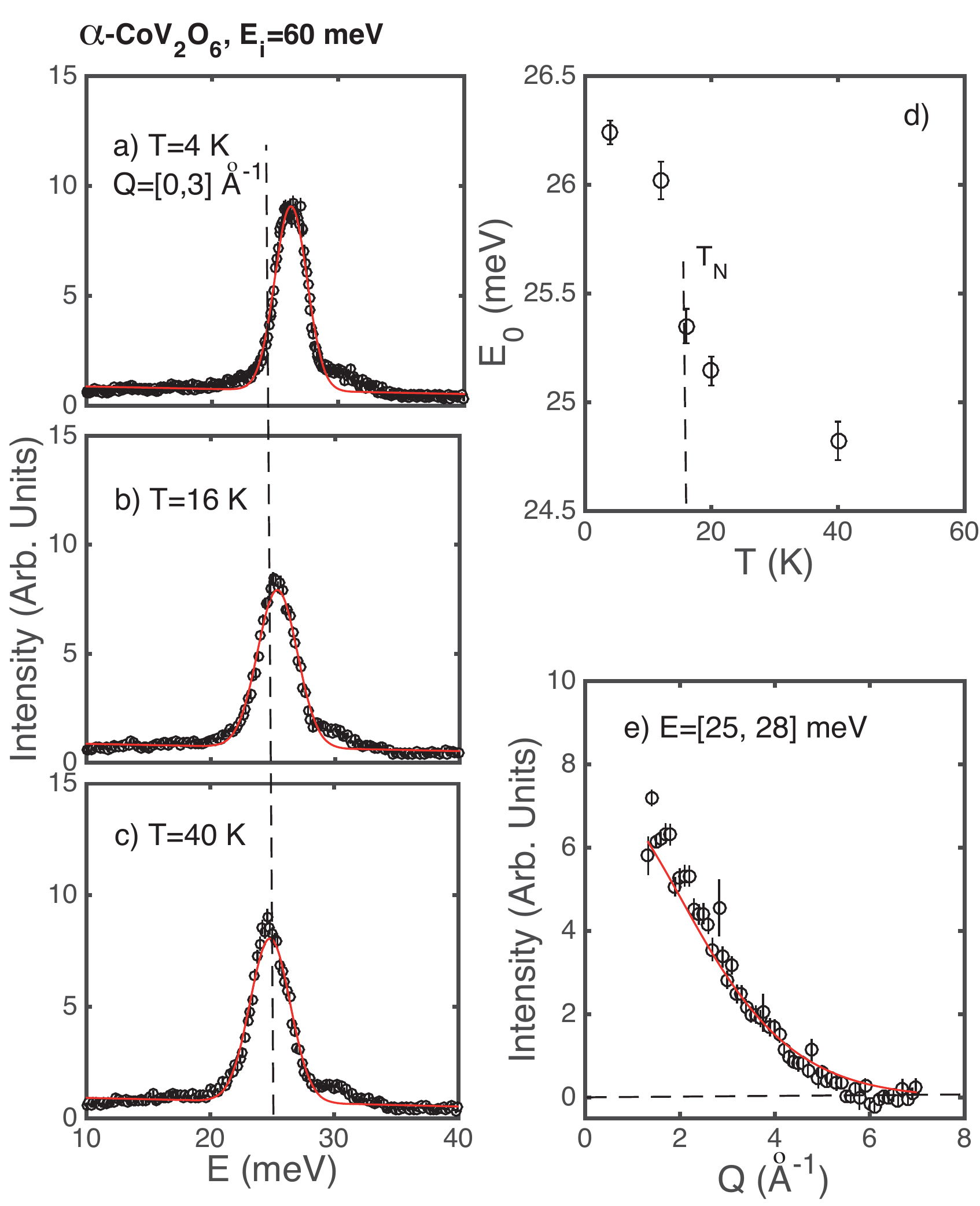}
\caption{\label{summary_60meV}  The energy, momentum and temperature dependence of the 24 meV excitation in monoclinic $\alpha$-CoV$_{2}$O$_{6}$.  $(a)$-$(c)$ shows the temperature dependence of this peak demonstrating little change in the energy structure with temperature.  $(d)$ shows a plot of the energy position as a function of temperature with an increase of $\sim$ 1 meV over the temperature range studied.  $(e)$ illustrates the momentum dependence at 24 meV with the solid curve the free ion Co$^{2+}$ form factor scaled to agree with the data.  We assign this transition to a transition from the lowest energy $j_{eff}={1\over 2}$ doublet to the first excited $j_{eff}={3\over 2}$ quartet.}
\end{figure}

\begin{figure}[t]
\includegraphics[width=8cm] {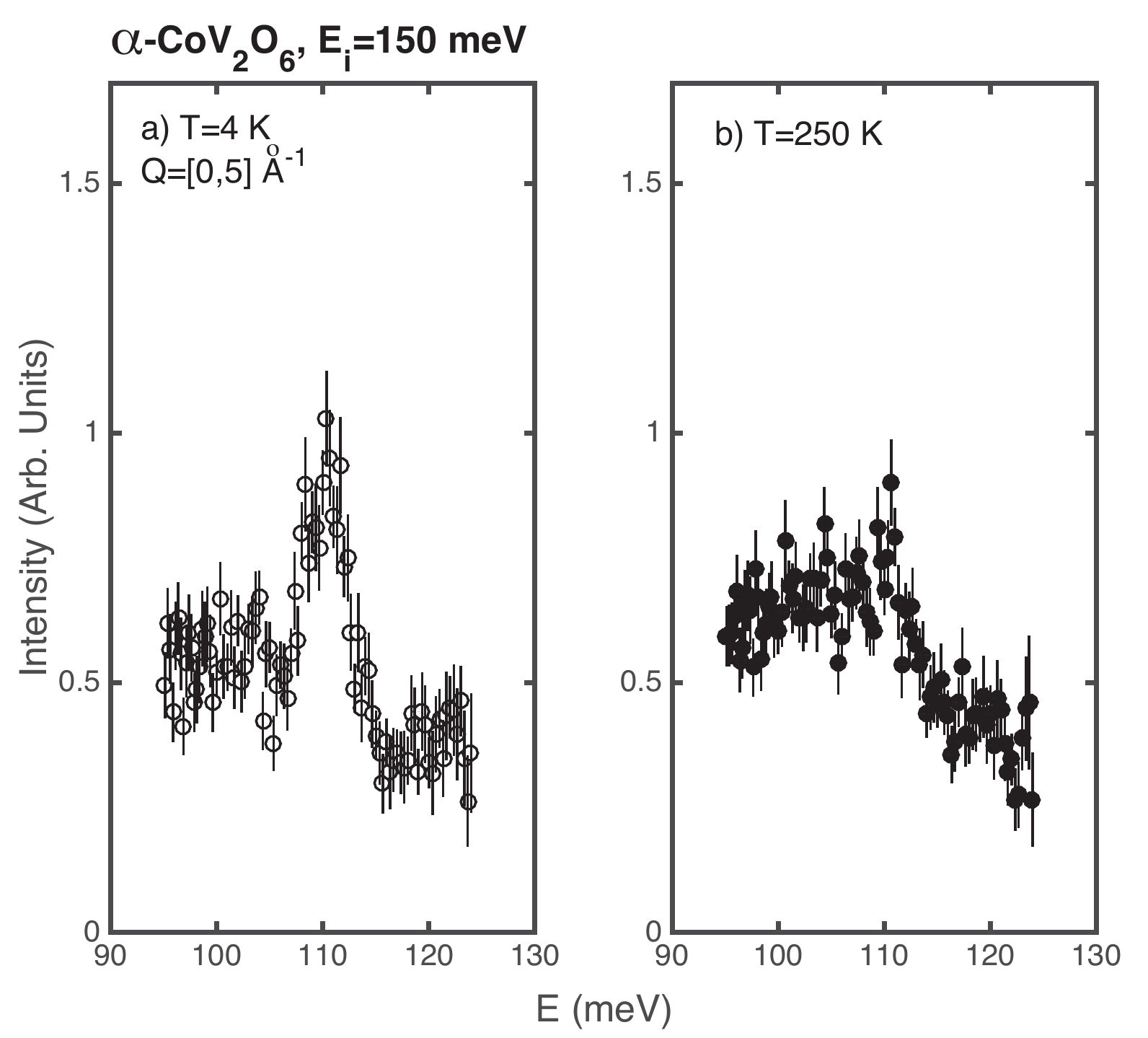}
\caption{\label{summary_150meV}  A plot of the 110 meV excitation in monoclinic $\alpha$-CoV$_{2}$O$_{6}$ at $(a)$ T=4 K and $(b)$ T=250 K.  The data show a change in intensity with temperature suggestive that this peak has a magnetic origin.  This transition is suggested to correspond to a transition to $j_{eff}={5\over2}$ multiplet.}
\end{figure}

\section{Results and Analysis}

Having outlined the essential single-ion theory (graphically illustrated in Fig. \ref{cef}), we now present neutron inelastic scattering results for powders of $\alpha$-monoclinic and $\gamma$-triclinic polymorphs.  The theory outlined above is only a single-ion model and does not include spin-exchange interactions and we now interpret the experimental neutron results in terms of this single-ion theory and also justify the neglect of further exchange terms in the Hamiltonian.  We consider the effects of spin exchange later in the paper.  We first consider the monolinic-$\alpha$ polymorph and then compare the results to the $\gamma$-triclinic variant.

\subsection{Monoclinic $\alpha$ polymorph (T$_{N}$=15 K)}

 A summary of the low temperature T=4 K results from MARI on the monoclinic polymorph are shown in Fig. \ref{mono_summary} for the three different incident energies (E$_{i}$=10, 60, and 150 meV) used.  These scans are taken in the magnetically ordered state.  The data show three peaks in the neutron response located at $\sim$ 4, 24, and 110 meV.  We focus our discussion on these excitations and return to the low-energy scattering present at $\sim$ 1 meV at the end of the paper.

Figure \ref{summary_10meV} illustrates the temperature dependence of the 4 meV peak.  Panel $(a)$ shows a constant momentum scan integrating over $Q=[0,1]$ \AA$^{-1}$ demonstrating that the peak is sharp in energy with the horizontal bar showing the calculated experiment energy resolution.  Panels $(b)$ and $(c)$ show the same scan but at 12 K (near T$_{N}$) and 40 K respectively.   Panels $(d)$ and $(e)$ show how the intensity and line width of this peak vary with temperature showing that it only appears as an underdamped peak below T$_{N}$.  The peak was fit to a lorentzian  ($I=I_{0}/(({E-E_{0})/ \Gamma}^{2}+1)$) convolved with the calculated resolution with a constant used to describe the background.  The intensity $I_{0}$ shows a clear drop at T$_{N}$ concomitant with broadening of the peak in energy indicating a strong dampening of the excitation.  These results clearly tie the presence of this excitation to magnetic order in $\alpha$-CoV$_{2}$O$_{6}$.

We therefore associate this peak at $\sim$ 4 meV with the splitting of the lowest energy ground state $j_{eff}={1\over 2}$ doublet owing to the molecular field induced by magnetic order.  This interpretation is based on the molecular field Zeeman splitting the doublet ground state.  While the result is similar to the inclusion of anisotropy terms in the Hamiltonian, this splitting and gapped excitation is the result of spin-orbit coupling in our Hamiltonian.  We note that the one-dimensional nature of the nuclear structure combined with the expected antiferromagnetic coupling (confirmed from magnetic neutron diffraction) of the spins ensures that the molecular field on each Co$^{2+}$ ion does not cancel in the magnetically ordered phase therefore creating a local nonzero $H_{mf}$ acting on each Co$^{2+}$ site at low temperatures.   

An important point to note is the resolution limited nature of this excitation within the ground state doublet which indicates little measurable magnetic coupling between the spins.  The effects of spin exchange on the powder averaged neutron cross section is discussed later in the paper and also presented in the Appendix in the context of the $\gamma$-triclinic polymorph.  The very weak exchange coupling between the Co$^{2+}$ spins indicate that any coupling introduced in the $H_{3}=H_{MF}S^{z}$ term of the single ion Hamiltonian discussed above are likely small in comparison to the expected spin-orbit coupling expected to be $|\lambda| =$ 16 meV from CoO.    We now discuss the higher energy excitations.

Figure \ref{summary_60meV} illustrates the temperature dependence and the momentum dependence of the 24 meV excitation measured on MARI with E$_{i}$=60 meV.  Panels $(a)$-$(c)$ show a constant momentum scans integrating in momentum over $Q=[0,3]$\AA$^{-1}$ at temperatures below T$_{N}$.  In contrast to the peak at 4 meV, the 24 meV excitation is present at all temperatures and does not display any broadening within experimental resolution.  We therefore associate this transition with a spin-orbit excitation from the $j_{eff}={1\over 2}$ doublet ground state to the higher energy $j={3\over 2}$ quartet.  This excitation is purely the result of the spin-orbit term in the Hamiltonian and is present both in the magnetically ordered and paramagnetic phases which is substantiated by the lack of any significant temperature dependence.  The observed energy scales is also similar to that measured in dilute (Mg,Co)O where this transition was found to be 34 meV.  

The peak position does show a small shift in energy which is associated with T$_{N}$ and this is shown in panel Fig. \ref{summary_60meV} $(d)$ which plots the peak position fitted from a single gaussian as a function of temperature.  The increase in energy of $\sim$ 1 meV is expected from the single-ion crystal field theory discussed above where magnetic order splits the lowest energy doublet and hence lowers the ground state energy therefore increasing the transition energy to other spin-orbit excitations (as illustrated in Fig. \ref{cef} c).

Figure \ref{summary_150meV} shows constant momentum scans taken with the high energy E$_{i}$=150 meV setting on MARI at T=4 K and 150 K.  The scans show a peak at $\sim$ 110 meV.  On heating, a change is observed in the line shape with a suppression at higher temperatures.  Given the high-energy scale and momentum dependence, we therefore associate this peak with a magnetic transition.  Based on the theory above, we associate this transition with excitations from the ground state to one of the members of the $j_{eff}=5/2$ multiplet.   Given these excitations, we now parametrize the magnetic scattering in terms of a single ion crystal field Hamiltonian.

\subsubsection{Crystal field ``model"}

To guide and substantiate the analysis above and the excitation assignments, we now parametrize the data in terms of a single ion Hamiltonian, calculating the transition energies and the neutron scattering intensities.

\begin{eqnarray}   
H=H_{SO}+H_{1}+H_{3}=\tilde{\lambda}\vec{l}\cdot\vec{S}+\Gamma_{z}l_{z}^{2}+H_{MF}S^{z}.
\label{cef_cub}
\end{eqnarray}

\noindent  To test whether this simplified single-ion Hamiltonian reproduces the energy scales, we have diagonalized this matrix fixing the spin-orbit coupling $\tilde{\lambda}$ to be 24 meV, the measured value in CoO (note that $\tilde{\lambda}\equiv -3/2 \lambda$, where $\lambda$ is defined in Ref. \onlinecite{Cowley13:88}).    We have taken this Hamiltonian acting as a perturbation on the ground state eigenstates of the cubic crystalline electric field $H_{0}$ discussed above and represented by the states $|l=1;s=3/2\rangle$.   We do not consider exchange terms between the Co$^{2+}$ sites given that the excitations are resolution limited in energy and show little sign of structure in momentum.

Given the main local octahedral distortion is an axial one (corresponding to $\Gamma_{z}$), we consider this term without an in-plane distortion.  The value of the parameter $\Gamma_{z}$ is unclear.  Motived by work in CoF$_{2}$, where the octahedra are flattened and $\Gamma_{z}$ was found to range from 1-2 $\times$ $\tilde{\lambda}$ (with the sign negative), we have considered the situation where $\Gamma_{z}$=-2$\lambda$.~\cite{Cowley73:6}  This single-ion Hamiltonian has transition energies of 28 , 44.5, and 114 meV with intensity ratios of 1, 0.15, and 0.02.   The experimental ratio of the 110 meV peak to the 24 meV peak is $\sim$ 0.06.    The single-ion model provides a good description of the 24 meV and the 110 meV excitations as spin-orbit transitions.  We have searched for intermediate excitations that may indicate the presence of  the 44.5 meV peak predicted from theory, however due to strong phonon scattering with increases with momentum transfer (Fig. \ref{mono_summary} $c)$) any magnetic transitions in this energy range are not discernible.  We note that the inclusion of a molecular field through $H_{MF}$ splits the ground state doublet introducing a gapped excitation.

\subsection{Triclinic $\gamma$ polymorph (T$_{N}$=7 K)}

\begin{figure}[t]
\includegraphics[width=8cm] {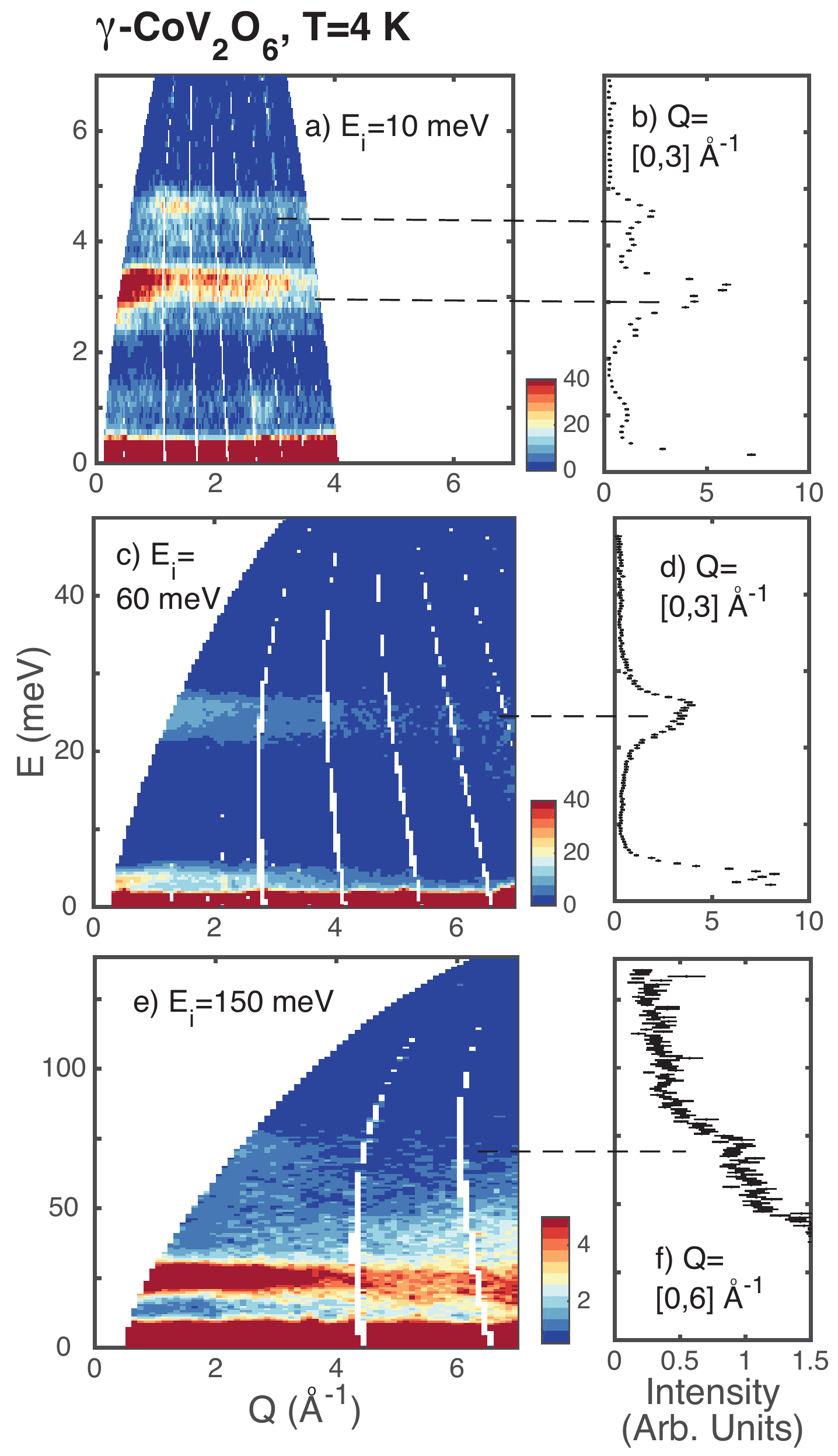}
\caption{\label{summary_tri}  A summary of the T=4 K neutron inelastic scattering measurements on the $\gamma$-CoV$_{2}$O$_{6}$  (triclinic polymorph) obtained on MARI with incident energies of E$_{i}$= $(a)$ 10, $(b)$ 60, and $(c)$ 150 meV.  All data was obtained in the magnetically ordered phase.}
\end{figure}

We now discuss the triclinic variant $\gamma$-CoV$_{2}$O$_{6}$ which has the additional complexity of having two different Co$^{2+}$ environments with proportions of  2:1.  A summary of the magnetic excitations in $\gamma$-CoV$_{2}$O$_{6}$ is illustrated in Fig. \ref{summary_tri} over the same dynamical ranges (with the same experimental configurations) as displayed previously for the $\alpha$-monoclinic polymorph in Fig. \ref{mono_summary}.  The summary shows a similar structure to the magnetic excitations as in the $\alpha$ (monoclinic) polymorph.   

Figure \ref{summary_tri} $(a)$ shows that the previous single sharp 4 meV is replaced by two bands of excitations at $\sim$ 3 and 5 meV.  The momentum dependence at small $Q$ displays a subtle dispersion to these bands of excitations giving structure in the peaks beyond the resolution determined by the spectrometer.  We discuss the two bands and the dispersion in terms of a 2-site model in the following section.

Figure \ref{summary_tri} $(b)$ displays a $\sim$ 24 meV peak at approximately the same energy as found in the $\alpha$ (monoclinic) polymorph.   This similarity further supports the assignment of this peak to a spin-orbit excitation from the ground state $j_{eff}=1/2$ to the $j_{eff}=3/2$ manifold of states.  Similar to the low-energy excitations reported in $(a)$ this peak is broadened in energy, again possibly the result of a weaker spin-spin exchange coupling.   At higher energies, the sharp 110 meV peak found in the monoclinic polymorph is not observable, but a band of scattering which decays with increasing momentum transfer is evident in panel $(c)$.  However, the observation of a weaker high-energy band is consistent with the $\alpha$ polymorph given the presence of two Co$^{2+}$ crystallographic sites and a different local octahedron distortion from the $\alpha$ polymorph (as noted in Ref. \onlinecite{Hollmann14:89}).  The single-ion model discussed above predicts a weaker transition in this energy range for a weaker axial distortion characterized by $\Gamma_{z}$.  In summary, we find the triclinic polymorph displays a similar energy structure to that of the $\alpha$ monoclinic polymorph but with broadened peaks consistent with larger spin-spin exchange and two crystallographic distinct Co$^{2+}$ sites which are less distorted than the $\alpha$-monoclinic polymorph.

There are two features that are different between $\alpha$-monclinic and $\gamma$-triclinic CoV$_{2}$O$_{6}$ which we consider within the single-ion framework discussed above.  First, the ordered moment obtained from neutron diffraction and magnetization is much larger in $\alpha$-monolinic polymorphs ($\sim$ 4.5 $\mu_{B}$) over the $\gamma$-triclinic variants ($\sim$ 3 $\mu_{B}$) interpreted as originating from a larger orbital contribution.  Second, the $\alpha$-monoclinic polymorph displays a weak $\sim$ 110 meV transition not observable in the $\gamma$-triclinic polymorph.  This excitation is nominally extinct in undistorted octahedra but is present due to additional terms in the Hamiltonian.  The combined results of a lower orbital contribution to the magnetic moment and the unobservable $\sim$ 110 meV excitation indicates a less distorted octahedra in $\gamma$-triclinic over $\alpha$-monoclinic.  This is consistent with the structural properties and can be seen by considering the local environment around the Co$^{2+}$ sites.  In particular, the deviation from a perfect octahedral environment is characterized by the following structural parameter. 

\begin{eqnarray}
\delta={1\over N} \sum_{i} \left({{d_{i}-\langle d \rangle} \over {\langle d \rangle}}\right)^{2} \times 10^{4} 
\label{cef_cub}
\end{eqnarray}

\noindent where $d_{i}$ are the distances from Co$^{2+}$ ion to the $N=6$ O$^{2-}$ in the distorted octahedra.  The average distance is denoted as $\langle d\rangle$.  The $\alpha$-monoclinic structure has $\delta=55$ and the $\gamma$-triclinic has $\delta=2.1$ and 4.8 for the two different Co$^{2+}$ sites.  The $\alpha$-monoclinic unit cell is therefore heavily distorted supporting a strong mixing substantiated by our neutron inelastic scattering study and the observation of an excitation from the $j_{eff}=1/2$ doublet ground state to the $j_{eff}=5/2$ manifold.    The $\gamma$ polymorph is comparatively less distorted implying less orbital mixing of the ordered moment and also the weakening of the $j_{eff}=1/2$ to $j_{eff}=5/2$ transition.   These results imply stronger orbital mixing in $\alpha$-monoclinic CoV$_{2}$O$_{6}$ over the triclinic polymorph.

The main driving term of this mixing in our heuristic model above for the single-ion Hamiltonian is the parameter $\Gamma_{z}$ which characterizes the axial distortion.  This term provides a means of coupling local strain to the magnetic properties through the spin-orbit coupling.  Because of the strong axial distortion, structural and magnetic properties are strongly coupled in CoV$_{2}$O$_{6}$ as seen through a large magnetostriction and dielectric anomalies discussed above.  This interpretation in terms of a single-ion theory is qualitatively in agreement with the conclusions derived from first principles calculations in Ref. \onlinecite{Kim12:85}.  CoV$_{2}$O$_{6}$ therefore represents a case where structural and magnetic orders are coupled through local crystalline electric fields.  We now return to the low-energy $\sim$ 4 meV excitations to understand the momentum dependence highlighted in Fig. \ref{summary_tri} (a) at low temperatures.

\begin{figure}[t]
\includegraphics[width=8.5cm] {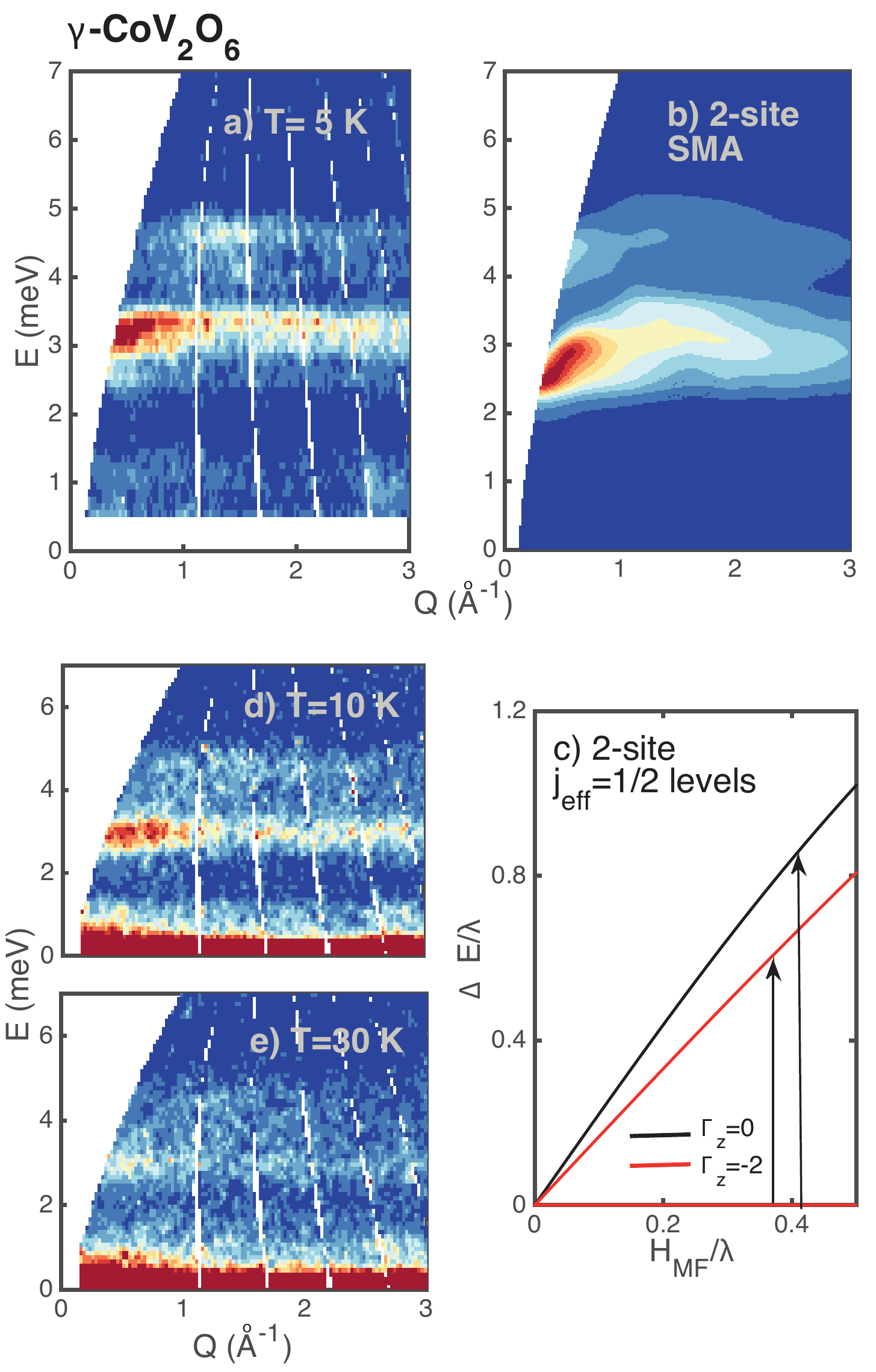}
\caption{\label{two_level}  $(a)$ Triclinic $\gamma$-CoV$_{2}$O$_{6}$ showing two bands of excitations and compared against the heuristic two-site model in $(b)$.  The effect of the differing octahedral distortions on the splitting of the ground state $j={1\over 2}$ doublet is illustrated in $(c)$ for \textit{isotropic} exchange in the absence of anisotropy (anisotropic exchange is discussed later in the text).  $(d-e)$ shows comparative scans above T$_{N}$.}
\end{figure}

\section{Anisotropy, exchange, and dimensionality}

In this section we discuss the exchange and anisotropy contributions to the Hamiltonian for both $\alpha$-monoclinic and $\gamma$-triclinic polymorphs of CoV$_{2}$O$_{6}$.  These terms are the most difficult to determine using powder averaged data, yet we discuss possible scenarios and apply sum rules of neutron scattering to extract information on the dimensionality of the interactions.  We also compare our results to other insulating Co$^{2+}$ containing materials.

\subsection{2-site $j_{eff}={1\over 2}$ model applied to $\gamma$-CoV$_{2}$O$_{6}$}

As identified above, a large difference between the magnetic excitations in $\alpha$-monoclinic and $\gamma$-triclinic polymorphs is the momentum and energy dependence of the low-energy excitations at $\sim$ 4 meV displayed in panels $(a)$ of Fig. \ref{summary_tri} and \ref{mono_summary}.   While monoclinic $\alpha$-CoV$_{2}$O$_{6}$ displays a sharp single dispersionless (within resolution limits) excitation with little observable momentum dependence in the powder average spectra, the $\gamma$ triclinic polymorph shows a contrasting strong momentum dependence and even a clear upward energy dispersion at small momentum transfers.  Another key difference between the two polymorphs is illustrated in Fig. \ref{two_level} $(d-e)$  which shows that the low-energy $\sim$ 5 meV excitation persists above T$_{N}$= 7 K and even up to 30 K in the triclinic $\gamma$-polymorph.  This contrasts with the picture presented for the $\alpha$ phase where the excitation rapidly decayed with temperatures above T$_{N}$.    Given the fact that both $\alpha,\gamma$ polymorphs display qualitatively similar bulk magnetic properties (magnetization plateaus) and are based upon a framework of octahedrally coordinated Co$^{2+}$, the underlying Hamiltonian describing the magnetism on the Co$^{2+}$ site must be similar.  

Previous studies (Ref. \onlinecite{Kimber11:84}) on the triclinic $\gamma$ polymorph have interpreted the excitations in terms of a ``soliton'' model originating from the underlying chain network in Fig. \ref{structure}. However, while that model may reproduce the ``ladder" of peaks in a constant momentum cut, it does not obviously carry over to provide a description of the monoclinic $\alpha$ polymorph where no such structure is observed in the energy dependence (Fig. \ref{mono_summary}).  A predominately one-dimensional model is also difficult to reconcile with the momentum dependence illustrated in Fig. \ref{summary_tri} which shows two connected levels at $\sim$ 4 meV and not a series of discrete excitations when scanned in energy.  The strongly one-dimensional picture of the CoV$_{2}$O$_{6}$ polymorphs is also contentious in the context of recent diffraction data which seem to suggest two-dimensional magnetism.~\cite{Lenertz11:115}

Given these contrasting results and in particular recent neutron magnetic diffraction experiments, we investigate whether an extension of the single-ion model discussed above in the context of the higher energy spin-orbit transitions can describe the low-energy physics consistently in both the $\alpha$ and $\gamma$ polymorphs.   In the dominant single-ion model discussed above, a molecular field is applied to each site in the magnetically ordered state with the strength determined by anisotropy and also exchange interactions between the spins.  This molecular field introduces a gap within the ground state doublet which is illustrated in Fig. \ref{two_level} (c) for two different sites with different axial distortions characterized by the parameter $\Gamma_{z}$.   Anisotropy terms due to the distorted octahedra and dipolar effects also contribute to the molecular field and will be different for the two different sites in the $\gamma$-triclinic polymorph given the differing local distortion.

To analyze the momentum dependence we first consider an additional symmetric Heisenberg type interaction in the model Hamiltonian discussed above with the form $H_{exchange}=\sum_{i,j} J_{ij} \vec{S}_{i}\cdot \vec{S}_{j}+H_{a}$, where $H_{a}$ includes anisotropy terms discussed above resulting from the heavily distorted oxygen octahedron surrounding the Co$^{2+}$ ion.  We later discuss anisotropic exchange and the possible experimental signatures of this in our data.  Given the large energy separating the ground state $j_{eff}={1\over 2}$ doublet from the first excited $j_{eff}={3\over 2}$ quartet ($\sim$ 300 K) and the fact that the fine structure is observed at low temperatures, we will only consider the effect of this additional term to the low-energy $j_{eff}={1\over 2}$ doublet.  

To test whether a 2-site model with weak exchange can describe the powder average data for $\gamma$-CoV$_{2}$O$_{6}$, we have performed a calculation powder averaging S($\vec{Q}$,E) with different gaps introduced through the differing molecular field contributions with the dispersion relation causing this discussed below.  A general form for the neutron scattering cross section can be derived from the first moment sum rule (Ref. \onlinecite{Hohenberg74:10}) which relates S($\vec{Q}$) to the dispersion $\epsilon(\vec{Q})$,

\begin{eqnarray}
S(\vec{Q})=-{2\over 3}{1\over \epsilon(\vec{Q})}\sum_{\vec{d}}J_{\vec{d}} \langle \vec{S_{0}} \cdot \vec{S_{\vec{d}}} \rangle [1-\cos(\vec{Q}\cdot\vec{d})].
\end{eqnarray}

\noindent This form for the cross section is independent of the specifics of the exchange interactions, but relies on the presence of an isotropic Heisenberg interaction in the Hamiltonian.  The first moment sum rule can be extended to include anisotropic interactions and the additional terms are discussed in Ref. \onlinecite{Zal05:book}.   However, these terms do not give a strong momentum dependence and therefore we have found that the powder averaged data is not strongly sensitive to these anisotropic terms in the sum rule above.  We therefore include only isotropic Heisenberg interactions for the first moment sum rule and discuss anisotropy $H_{a}$ below in the context of the dispersion relation $\epsilon$.  Further discussion of anisotropic exchange is presented in the next section in the context of an additional low-energy mode observed in both structural polymorphs.

In the single mode approximation where only one transition level is being considered (for example when we are within the lowest energy manifold of states), the measured structure factor can be written in terms of a momentum dependent part and a Dirac delta function forcing energy conservation $S(\vec{Q},E)=S(\vec{Q})\delta(E-\epsilon(\vec{Q}))$.  For numerical purposes, we have approximated $\delta(E)$ as a Lorentzian, with a full-width equal to the calculated experimental resolution width on MARI (given above in the experimental details section).   To account for the dispersion, we have used a heuristic form which obeys the lattice periodicity - $\epsilon(\vec{Q})=\beta_{0}+\sum_{i} \beta_{i} \cos(\vec{Q}\cdot \vec{d_{i}})$ where $d_{i}$ are vectors connecting the Co$^{2+}$ ions to the nearest neighbors and $\beta_{0}$ represents the gap energy (characteristic of the splitting described in Fig. \ref{two_level} $c$) and $\beta_{i}$ are related to the strength of the exchange to the nearest neighbors.   We note that a similar approach has been applied to other low dimensional magnets and a further description of this analysis can be found elsewhere.~\cite{Stock09:103,Stock12:85,Xu00:84,Stone01:64,Hammar98:57}  

While exact forms for the energy dispersion can be calculated (for example see Ref. \onlinecite{Pfeuty70:57}), given the ambiguity from powder averaging and the loss of information regarding anisotropic exchange, we have chosen to use this heuristic expansion which obeys the symmetry of the lattice.  Physically, the excitation energy gap $\beta_{0}$ can be related to the anisotropy and the gap introduced through splitting the low-energy doublet when projecting from a full $S={3\over2}$ description to a $j_{eff}={1\over 2}$ ground state doublet.  Expressions for the transformation including different terms due to anisotropy are given in Ref. \onlinecite{Buyers86:56} for tetragonal KCoF$_{3}$.  This term incorporates the local molecular field and anisotropy (Ref. \onlinecite{Schron12:86}) and therefore should be different for the two different Co$^{2+}$ sites with different local distortions in the $\gamma$-triclinic variant.     As discussed in Ref. \onlinecite{Cabrera14:90}, the parameters $\beta_{i}$ can be interpreted as hopping terms along different bond directions where single spin-flips cost energy $t \sim SJ$.   

The calculation for 2-sites with volume proportions of 2:1 is shown in Fig. \ref{two_level} $(b)$ and compared against experimental data in $(a)$.  For this calculation, we have considered two different dispersions with differing values of $\beta_{0}$ to model the different local distortions.   We have then calculated the sum over the two sites $S_{tot}= \nu_{2} S_{Co1}+ \nu_{2} S_{Co2}$ with the relative weight of the two excitations $\nu_{1}:\nu_{2}$  being 2:1 following the volume proportions.  Interestingly, to get the distribution of intensity shown, particularly at small momentum transfers, we have needed to input two ``hopping" parameters $\beta_{i}$ connecting spins both along the chains and also between the chains with $\beta_{1}/\beta_{2}$=0.75 and $J_{\vec{a}}\langle \vec{S}_{0}\cdot \vec{S}_{a} \rangle/J_{\vec{b}}\langle \vec{S}_{0}\cdot \vec{S}_{b} \rangle$=1.   The powder data is therefore suggestive of two-dimensional model in the triclinic ($\gamma$) polymorph instead of a strongly one-dimensional.  The intensity ratio of the two modes at 4 and 5 meV, and the absence of such splitting in the $\alpha$-monclinic variant, leads us to suggest that the fine structure and the two levels originates from the two different crystallographic sites in $\gamma$-triclinic CoV$_{2}$O$_{6}$.  While a single crystal analysis is required for fully conclusive statements, the heuristic model described here seems to imply that $\gamma$-CoV$_{2}$O$_{6}$ consists of ferromagnetically coupled planes consistent with diffraction studies.~\cite{Lenertz11:115}  The two sites in the $\gamma$-triclinic polymorph therefore not only introduces two different sites, but also seems to break up the chains into a two dimensional framework.  Given the lack of dispersion or momentum dependence in the $\alpha$-monoclinic variant, we are not able to make statements on the dimensionality of the exchange interaction in that system based on our powder data.

Another difference between $\alpha$ and $\gamma$ is the temperature dependence of the $\sim$ 4 meV excitation which persists to very high temperatures in the triclinic $\gamma$ polymorph (shown in Figs. \ref{two_level} d-e), but not the $\alpha$ monoclinic variant discussed above.   Figs. \ref{two_level} also seems to illustrate the strong momentum dependence, particularly at small momentum transfers, is lost at higher temperatures.  These combined results can be understood in terms of the single-ion model presented above in the presence of two-dimensional interactions.  In terms of the single-ion picture described above, a local field on an individual Co$^{2+}$ can remain above the magnetic ordering temperature provided there is a finite correlation length of ordered spins.  While there have been no theoretical calculations performed for the correlation length on the CoV$_{2}$O$_{6}$ lattice studied here, for a two dimensional square lattice with $S={1\over 2}$, as in La$_{2}$CuO$_{4}$ and Sr$_{2}$CuO$_{2}$Cl$_{2}$ (Refs. \onlinecite{Birgeneau95;56,Greven95:96}), it has been shown that $\xi \propto \exp(1.25 J/kT)$ in agreement with simulations (Ref. \onlinecite{Makivic91:43}) and also the 2D-nonlinear sigma model (Ref. \onlinecite{Hasenfratz83:50}).  This relation implies that the stronger $J$, the larger the correlation length is at higher temperatures.  In the context of the $\alpha,\gamma$-CoV$_{2}$O$_{6}$, this implies that the correlation length would fall off more quickly with increasing temperature in the monoclinic $\alpha$ variant in comparison to the $\gamma$ triclinic variant with the large exchange.  The persistence of the $\sim$ 4 meV peak in the $\gamma$-triclinic polymorph and not in the $\alpha$-monoclinic polymorph is therefore consistent with stronger exchange in the triclinic sample and also two-dimensional interactions suggested by the single-mode analysis described above.  

In summary, we have applied the first sum rule and found that a heuristic two-site model describes the $\gamma$-triclinic variant with two-dimensional interactions.  One thing that is lacking in this heuristic description is a description of the anisotropy term in the Hamiltonian (referred to as $H_{a}$ above) resulting from dipolar effects and also an anisotropy in the exchange constants.  Given the large local distortion noted above for both the $\alpha$-monolinic and $\gamma$-triclinic variant and the strong mixing of the various spin-orbit levels, such additional terms are expected in the Hamiltonian.    An experimental signature of this anisotropy is the breaking of the degeneracy of the low-temperature spin-waves and we now apply high resolution neutron spectroscopy to search for this in the magnetically ordered state in both polymorphs at low energies.  


\subsection{Low-energy excitations and possible signatures of anisotropic exchange}

\begin{figure}[t]
\includegraphics[width=8.5cm] {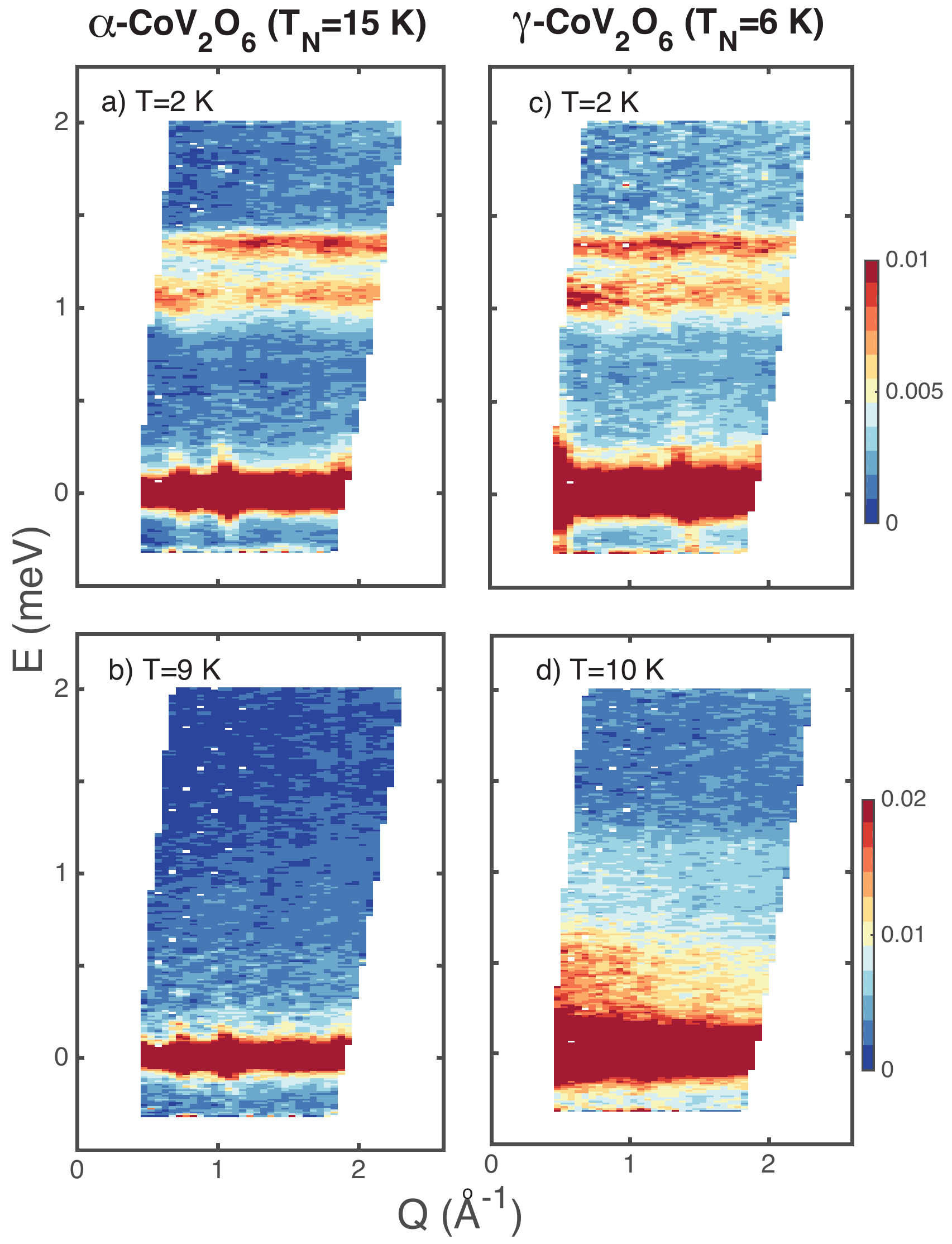}
\caption{\label{IRIS_summary}  A comparison of the low-energy excitations at $\sim$ 1 meV in both the $\alpha$-monoclinic ($a$ and $b$) and $\gamma$-triclinic ($c$ and $d$) polymorphs.  The excitations appear at the same energy in both polymorphs and disappear rapidly with temperature.}
\end{figure}

The above discussion has focussed on the single-ion effects and found a strong response of the excitations to the local crystal field supporting the notion that a strongly distorted octahedra causes stronger orbital mixing in the $\alpha$ polymorph over the $\gamma$ material.  All inelastic transitions were well accounted for in terms of the single-ion model presented in the introduction with the $\gamma$ polymorph also requiring a two-dimensional interactions modelled heuristically through the combined use of the first moment sum rule and a ``hopping" model to write down an energy dispersion.  However, we note that all of the magnetic excitations discussed above either are present above T$_{N}$ or decay at T$_{N}$.  This makes them unlikely to be directly tied to the magnetization plateaus which disappear quickly with temperature, and at least in the monoclinic $\alpha$ polymorph, well below T$_{N}$.  One unresolved aspect of the single-ion model was that the calculation predicts a large dominant intensity for excitations within the $j_{eff}=1/2$ ground state doublet. As seen in Fig. \ref{mono_summary}, this is inconsistent with a comparison between the $\sim$ 24 meV and $\sim$ 4 meV excitation (see Fig. \ref{mono_summary} (b) for comparison on the same scale with the same experimental configuration).    

This represents a breakdown of the dominant single-ion picture for a full description of the excitations and indicates the presence of spectral weight elsewhere not discussed above.  Another low-energy excitation is present and can be seen in both Fig. \ref{mono_summary} $(a)$ and Fig. \ref{summary_tri} $(a)$ at an energy transfer of $\sim$ 1 meV.  We further investigated this excitation using the IRIS backscattering spectrometer which offers high energy resolution of 17.5 $\mu eV$ at low-energy transfers.  Fig. \ref{IRIS_summary} shows momentum and energy slices taken on IRIS comparing the results for the $\alpha$-monoclinic and $\gamma$-triclinic polymorphs.  A strong band of magnetic excitations which consists of two peaks in energy is seen at the same energy in both the $\alpha$-monoclinic and $\gamma$-triclinic polymorphs and decays rapidly with temperature.   The energy scale of the band is similar in both $\alpha$ monoclinic and $\gamma$ triclinic polymorphs despite different local distortions discussed above.

Given the strong local distortion in the local octahedra in both $\alpha,\gamma$ polymorphs, we expect anisotropy terms in the Hamiltonian and in particular anisotropic exchange between the Co$^{2+}$ ions.  The presence of anisotropic exchange would modify the exchange terms to the magnetic Hamiltonian.  Such a modification was discussed in the context of tetragonal KCoF$_{3}$ (Ref. \onlinecite{Buyers86:56}) and was written as follows for two interacting spins (labelled as 1 and 2),

\begin{eqnarray}
H_{ex}^{12}=J_{\parallel}S_{z}^{1}S_{z}^{2}+J_{\perp}(S_{x}^{1}S_{x}^{2}+S_{y}^{1}S_{y}^{2}).
\end{eqnarray}

\noindent Such anisotropy has even been suggested to exist in comparatively undistorted octahedra (Ref. \onlinecite{Elliott68:39}) and an experimental signature of this is the splitting of the magnon branches in the magnetically ordered phase.  The effects of anisotropy were discussed in the context of tetragonal KCoF$_{3}$ where it was noted that the splitting of the degeneracy scaled as ${\Delta E \over E}=3(J_{\parallel}-J_{\perp})/4J_{\perp}$ (with $J_{\perp}$ and $J_{\parallel}$ being to the two distinguishing directions).   The case of a strong anisotropic exchange may explain the existence of two branches to the low-energy fluctuations which are linked with the magnetically ordered phase.  The two modes are present even in monoclinic phase where there is only one Co$^{2+}$ site.  

To corroborate this picture we have searched for other compounds which have a similar local bonding environment to that of $\alpha,\gamma$-CoV$_{2}$O$_{6}$.  Co$_{3}$V$_{2}$O$_{8}$ has a similar local framework based on a distorted octahedra, but two different Co$^{2+}$ sites giving a ``distorted-Kagome" type magnetic lattice which also displays magnetic field induced transitions.~\cite{Wilson07:19,Wilson07:75,Szymczak06:73,Petrenko10:82,Fritsch12:86,Helton12:24}  Single crystal work on that compound has revealed also two low-energy branches with similar energy scales to that observed here for CoV$_{2}$O$_{6}$.~\cite{Rama09:79}  The similarity in the energy scales indicates a predominately anisotropic model may provide a consistent description.  Further work on single crystals in CoV$_{2}$O$_{6}$ and other materials based on a similar framework extracting the polarization of the magnetic fluctuations will ultimately aid in linking these systems and also resolving these lower energy components to the Hamiltonian and determining their possible role in magnetic field induced transitions or magnetization plateaus.

A contrasting feature of these low-energy fluctuations in comparison to the higher energy excitations discussed above is shown in Fig \ref{IRIS_summary} which show the low-energy $\sim$ 1 meV fluctuations in both $\alpha$ and $\gamma$ polymorphs decay rapidly with temperature.  This is particular true for $\alpha$-CoV$_{2}$O$_{6}$ where the excitations are absent at 9 K, well below T$_{N}$=15 K for this compound.   The magnetic excitations also have fine structure which is only resolved because of the high resolution on IRIS.  From the current powdered average data set, it is not possible to determine if these are two separate excitations or one single dispersing band.   We note that for triclinic $\gamma$-CoV$_{2}$O$_{6}$, at T=10 K the two bands are largely replaced by quasielastic scattering near Q=0 implying dominant ferromagnetic coupling, consistent with the single mode analysis presented above.  

It is clear from this analysis that the excitations within the $j_{eff}=1/2$ doublet are divided into two bands with one decaying at T$_{N}$ ($\sim$ 4 meV) and the other at much lower energies and also much more sensitive to temperature ($\sim$ 1 meV).   The correlation with temperature between the low-energy band and the plateaus in the magnetization may indicate that splitting of magnon degeneracy through anisotropic exchange to be central for facilitating steps in the magnetization.  The anisotropic exchange is directly tied with the distorted octahedra making this consistent with first principles calculations in Ref. \onlinecite{Kim12:85} which imply that a dominant single ion scenario maybe enough for magnetization plateaus.

\section{Discussion and Conclusions}

We have presented a study of the magnetic excitations in $\alpha$-monoclinic and $\gamma$-triclinic polymorphs of CoV$_{2}$O$_{6}$ and interpreted them in terms of a dominant single-ion Hamiltonian.  We have found well separated magnetic excitations in both compounds consistent with spin-orbit excitations from a ground state $j_{eff}=1/2$ doublet to excited $j_{eff}=3/2$ and $j_{eff}=5/2$ multiplets.  The energy positions are set by the spin-orbit coupling and also a dominant axial distortion of the local octahedra.  The highly distorted octahedron in $\alpha$-CoV$_{2}$O$_{6}$ supports stronger orbital mixing resulting in allowed neutron transitions from the $j_{eff}=1/2$ ground state to the excited $j_{eff}=5/2$ level.  It is also consistent with a larger orbital moment observed with x-rays and suggested by magnetization studies.  

The sharp excitations combined with weak momentum dependence show that any spin exchange is much weaker than the spin-orbit coupling $\tilde{\lambda}=\alpha \lambda$=24 meV.  In the $\alpha$-monoclinic polymorph, the exchange is very weak indicated by the presence of resolution limited in energy excitations with little dispersion in momentum.  The $\gamma$-triclinic polymorph shows energy broadened excitations and therefore evidence of stronger spin exchange, however still significantly less than the dominant spin-orbit coupling.  A heuristic model using the first moment sum rule implies that the spin coupling in the $\gamma$-triclinic variant is predominately two dimensional consistent with current magnetic diffraction studies.   This analysis relies on the presence of a gap and the distribution of intensity within the weakly dispersing band.   

While these single ion excitations are not directly associated with the plateaus, we also report bands of excitations at $\sim$ 1 meV that exists at the same energy in both the $\alpha$-monoclinic and $\gamma$-triclinic polymorphs and decay in intensity well below T$_{N}$ in the $\alpha$-monolinic material.   While the origin of these excitations is not clear from the powdered average data, the temperature dependence is suggestive that these excitations maybe connected with the magnetization plateaus.  We suggest that these excitations originate from strong anisotropic exchange and have compared them to available single crystal data in other low-dimensional Co$^{2+}$ compounds.  

\begin{figure}[t]
\includegraphics[width=8.2cm] {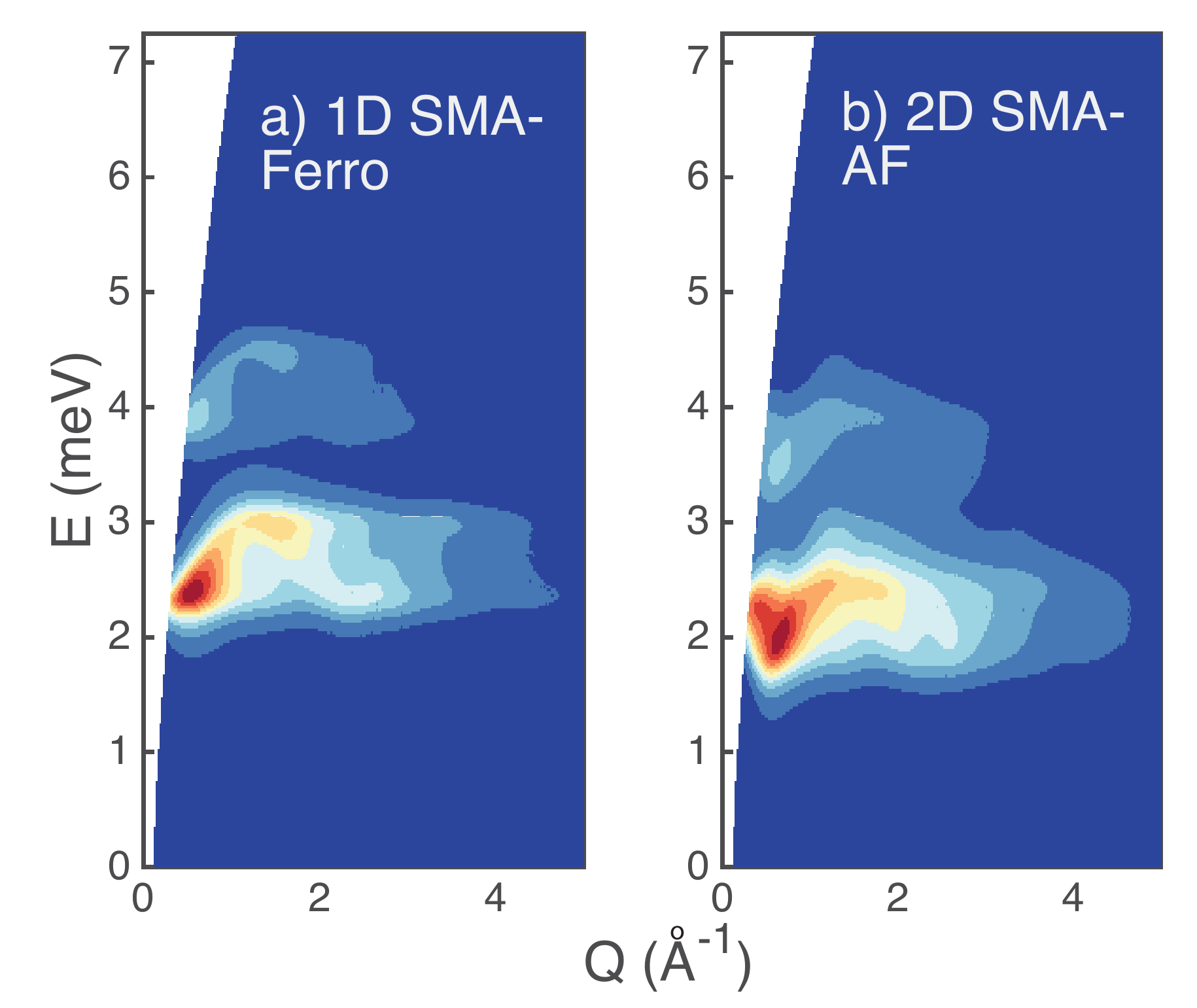}
\caption{\label{compare_calc}  Calculations within the 2-site $j_{eff}={1\over 2}$ model (in the context of the $\gamma$-triclinic polymorph) discussed in the appendix and compared against data and calculations presented in the main text. $(a)$ is is a calculation for ferromagnet uncoupled chains. The concentration of spectral weight at low energies and momentum transfers prefers another description.  $(b)$ is the a calculation for a two dimensional interaction.  The concentration of spectral weight at finite momentum transfers excludes this description of the data.}
\end{figure}

\section{Acknowledgements}

We are grateful to the Carnegie Trust for the Universities of Scotland, the Royal Society, STFC, and the EPSRC for support of this work.  Information regarding open access to data is provided in the supplementary information.~\cite{supp}

\section{Appendix}

In the discussion above, we applied a heuristic calculation using the single mode approximation to study the momentum dependence of the low-energy spin fluctuations.  Further details of this calculation and comparison of various models is shown in this Appendix.  

While the powder averaging results in a loss of information about exchange anisotropy, the presence of a gap in the excitation spectrum results in a sensitivity to the dimensionality of the exchange interactions.  The momentum dependence of the powder averaged intensity also provides a sensitivity to the sign of the exchange interaction.  We demonstrate this by showing in Fig. \ref{compare_calc} two further calculations with a strictly one-dimensional model (panel $a$) and also a two-dimensional antiferromagnetic model in panel $(b)$.   We have favored a two-dimensional ferromagnetic model in the main text over the one-dimensional model as the 1D calculation concentrates the spectral at lower energies which is not consistent with data presented above.  This general feature was found to be independent of the value of the exchange constant and a property of the dimensionality.  The antiferromagnetic model in panel $(b)$ shows that the spectral weight is concentrated at finite momentum transfers which is inconsistent with the data.  While we emphasize that a single crystal experiment is required to conclusively derive the interactions and the dimensionality, the powder averaged inelastic response does seem to imply a two-dimensional ferromagnetic model is preferred. 
 

\begin{thebibliography}{80}%
\makeatletter
\providecommand \@ifxundefined [1]{%
 \@ifx{#1\undefined}
}%
\providecommand \@ifnum [1]{%
 \ifnum #1\expandafter \@firstoftwo
 \else \expandafter \@secondoftwo
 \fi
}%
\providecommand \@ifx [1]{%
 \ifx #1\expandafter \@firstoftwo
 \else \expandafter \@secondoftwo
 \fi
}%
\providecommand \natexlab [1]{#1}%
\providecommand \enquote  [1]{``#1''}%
\providecommand \bibnamefont  [1]{#1}%
\providecommand \bibfnamefont [1]{#1}%
\providecommand \citenamefont [1]{#1}%
\providecommand \href@noop [0]{\@secondoftwo}%
\providecommand \href [0]{\begingroup \@sanitize@url \@href}%
\providecommand \@href[1]{\@@startlink{#1}\@@href}%
\providecommand \@@href[1]{\endgroup#1\@@endlink}%
\providecommand \@sanitize@url [0]{\catcode `\\12\catcode `\$12\catcode
  `\&12\catcode `\#12\catcode `\^12\catcode `\_12\catcode `\%12\relax}%
\providecommand \@@startlink[1]{}%
\providecommand \@@endlink[0]{}%
\providecommand \url  [0]{\begingroup\@sanitize@url \@url }%
\providecommand \@url [1]{\endgroup\@href {#1}{\urlprefix }}%
\providecommand \urlprefix  [0]{URL }%
\providecommand \Eprint [0]{\href }%
\providecommand \doibase [0]{http://dx.doi.org/}%
\providecommand \selectlanguage [0]{\@gobble}%
\providecommand \bibinfo  [0]{\@secondoftwo}%
\providecommand \bibfield  [0]{\@secondoftwo}%
\providecommand \translation [1]{[#1]}%
\providecommand \BibitemOpen [0]{}%
\providecommand \bibitemStop [0]{}%
\providecommand \bibitemNoStop [0]{.\EOS\space}%
\providecommand \EOS [0]{\spacefactor3000\relax}%
\providecommand \BibitemShut  [1]{\csname bibitem#1\endcsname}%
\let\auto@bib@innerbib\@empty
\bibitem [{\citenamefont {Collins}\ and\ \citenamefont
  {Petrenko}(1997)}]{Collins97:75}%
  \BibitemOpen
  \bibfield  {author} {\bibinfo {author} {\bibfnamefont {M.~F.}\ \bibnamefont
  {Collins}}\ and\ \bibinfo {author} {\bibfnamefont {O.~A.}\ \bibnamefont
  {Petrenko}},\ }\href@noop {} {\bibfield  {journal} {\bibinfo  {journal} {Can.
  J. Phys.}\ }\textbf {\bibinfo {volume} {75}},\ \bibinfo {pages} {605}
  (\bibinfo {year} {1997})}\BibitemShut {NoStop}%
\bibitem [{\citenamefont {Stock}\ \emph {et~al.}(2009)\citenamefont {Stock},
  \citenamefont {Chapon}, \citenamefont {Adamopoulos}, \citenamefont {Lappas},
  \citenamefont {Giot}, \citenamefont {Taylor}, \citenamefont {Green},
  \citenamefont {Brown},\ and\ \citenamefont {Radaelli}}]{Stock09:103}%
  \BibitemOpen
  \bibfield  {author} {\bibinfo {author} {\bibfnamefont {C.}~\bibnamefont
  {Stock}}, \bibinfo {author} {\bibfnamefont {L.~C.}\ \bibnamefont {Chapon}},
  \bibinfo {author} {\bibfnamefont {O.}~\bibnamefont {Adamopoulos}}, \bibinfo
  {author} {\bibfnamefont {A.}~\bibnamefont {Lappas}}, \bibinfo {author}
  {\bibfnamefont {M.}~\bibnamefont {Giot}}, \bibinfo {author} {\bibfnamefont
  {J.~W.}\ \bibnamefont {Taylor}}, \bibinfo {author} {\bibfnamefont {M.~A.}\
  \bibnamefont {Green}}, \bibinfo {author} {\bibfnamefont {C.~M.}\ \bibnamefont
  {Brown}}, \ and\ \bibinfo {author} {\bibfnamefont {P.~G.}\ \bibnamefont
  {Radaelli}},\ }\href@noop {} {\bibfield  {journal} {\bibinfo  {journal}
  {Phys. Rev. Lett.}\ }\textbf {\bibinfo {volume} {103}},\ \bibinfo {pages}
  {077202} (\bibinfo {year} {2009})}\BibitemShut {NoStop}%
\bibitem [{\citenamefont {Tennant}\ \emph {et~al.}(1993)\citenamefont
  {Tennant}, \citenamefont {Perring}, \citenamefont {Cowley},\ and\
  \citenamefont {Nagler}}]{Tennant93:73}%
  \BibitemOpen
  \bibfield  {author} {\bibinfo {author} {\bibfnamefont {D.~A.}\ \bibnamefont
  {Tennant}}, \bibinfo {author} {\bibfnamefont {T.~G.}\ \bibnamefont
  {Perring}}, \bibinfo {author} {\bibfnamefont {R.~A.}\ \bibnamefont {Cowley}},
  \ and\ \bibinfo {author} {\bibfnamefont {S.~E.}\ \bibnamefont {Nagler}},\
  }\href@noop {} {\bibfield  {journal} {\bibinfo  {journal} {Phys. Rev. Lett.}\
  }\textbf {\bibinfo {volume} {70}},\ \bibinfo {pages} {4003} (\bibinfo {year}
  {1993})}\BibitemShut {NoStop}%
\bibitem [{\citenamefont {Lake}\ \emph {et~al.}(2005)\citenamefont {Lake},
  \citenamefont {Tennant}, \citenamefont {Frost},\ and\ \citenamefont
  {Nagler}}]{Lake05:4}%
  \BibitemOpen
  \bibfield  {author} {\bibinfo {author} {\bibfnamefont {B.}~\bibnamefont
  {Lake}}, \bibinfo {author} {\bibfnamefont {D.~A.}\ \bibnamefont {Tennant}},
  \bibinfo {author} {\bibfnamefont {C.~D.}\ \bibnamefont {Frost}}, \ and\
  \bibinfo {author} {\bibfnamefont {S.~E.}\ \bibnamefont {Nagler}},\
  }\href@noop {} {\bibfield  {journal} {\bibinfo  {journal} {Nat. Mater.}\
  }\textbf {\bibinfo {volume} {4}},\ \bibinfo {pages} {329} (\bibinfo {year}
  {2005})}\BibitemShut {NoStop}%
\bibitem [{\citenamefont {Piazza}\ \emph {et~al.}(2014)\citenamefont {Piazza},
  \citenamefont {Mourigal}, \citenamefont {Nilsen}, \citenamefont
  {Tregenna-Piggott}, \citenamefont {Perring}, \citenamefont {Enderle},
  \citenamefont {McMorrow}, \citenamefont {Ivanov},\ and\ \citenamefont
  {Ronnow}}]{Piazza14:11}%
  \BibitemOpen
  \bibfield  {author} {\bibinfo {author} {\bibfnamefont {B.~D.}\ \bibnamefont
  {Piazza}}, \bibinfo {author} {\bibfnamefont {M.}~\bibnamefont {Mourigal}},
  \bibinfo {author} {\bibfnamefont {N.~B. C. G.~J.}\ \bibnamefont {Nilsen}},
  \bibinfo {author} {\bibfnamefont {P.}~\bibnamefont {Tregenna-Piggott}},
  \bibinfo {author} {\bibfnamefont {T.~G.}\ \bibnamefont {Perring}}, \bibinfo
  {author} {\bibfnamefont {M.}~\bibnamefont {Enderle}}, \bibinfo {author}
  {\bibfnamefont {D.~F.}\ \bibnamefont {McMorrow}}, \bibinfo {author}
  {\bibfnamefont {D.~A.}\ \bibnamefont {Ivanov}}, \ and\ \bibinfo {author}
  {\bibfnamefont {H.~M.}\ \bibnamefont {Ronnow}},\ }\href@noop {} {\bibfield
  {journal} {\bibinfo  {journal} {Nat. Phys.}\ }\textbf {\bibinfo {volume}
  {11}},\ \bibinfo {pages} {62} (\bibinfo {year} {2014})}\BibitemShut {NoStop}%
\bibitem [{\citenamefont {Haldane}(1983)}]{Haldane83:50}%
  \BibitemOpen
  \bibfield  {author} {\bibinfo {author} {\bibfnamefont {F.~D.~M.}\
  \bibnamefont {Haldane}},\ }\href@noop {} {\bibfield  {journal} {\bibinfo
  {journal} {Phys. Rev. Lett.}\ }\textbf {\bibinfo {volume} {50}},\ \bibinfo
  {pages} {1153} (\bibinfo {year} {1983})}\BibitemShut {NoStop}%
\bibitem [{\citenamefont {Buyers}\ \emph {et~al.}(1971)\citenamefont {Buyers},
  \citenamefont {Holden}, \citenamefont {Svensson}, \citenamefont {Cowley},\
  and\ \citenamefont {Hutchings}}]{Buyers86:56}%
  \BibitemOpen
  \bibfield  {author} {\bibinfo {author} {\bibfnamefont {W.~J.~L.}\
  \bibnamefont {Buyers}}, \bibinfo {author} {\bibfnamefont {T.~M.}\
  \bibnamefont {Holden}}, \bibinfo {author} {\bibfnamefont {E.~C.}\
  \bibnamefont {Svensson}}, \bibinfo {author} {\bibfnamefont {R.~A.}\
  \bibnamefont {Cowley}}, \ and\ \bibinfo {author} {\bibfnamefont {M.~T.}\
  \bibnamefont {Hutchings}},\ }\href@noop {} {\bibfield  {journal} {\bibinfo
  {journal} {J. Phys. C: Solid St. Phys.}\ }\textbf {\bibinfo {volume} {4}},\
  \bibinfo {pages} {2139} (\bibinfo {year} {1971})}\BibitemShut {NoStop}%
\bibitem [{\citenamefont {Kenzelmann}\ \emph {et~al.}(2002)\citenamefont
  {Kenzelmann}, \citenamefont {Cowley}, \citenamefont {Buyers}, \citenamefont
  {Tun}, \citenamefont {Coldea},\ and\ \citenamefont
  {Enderle}}]{Kenzelmann02:66}%
  \BibitemOpen
  \bibfield  {author} {\bibinfo {author} {\bibfnamefont {M.}~\bibnamefont
  {Kenzelmann}}, \bibinfo {author} {\bibfnamefont {R.~A.}\ \bibnamefont
  {Cowley}}, \bibinfo {author} {\bibfnamefont {W.~J.~L.}\ \bibnamefont
  {Buyers}}, \bibinfo {author} {\bibfnamefont {Z.}~\bibnamefont {Tun}},
  \bibinfo {author} {\bibfnamefont {R.}~\bibnamefont {Coldea}}, \ and\ \bibinfo
  {author} {\bibfnamefont {M.}~\bibnamefont {Enderle}},\ }\href@noop {}
  {\bibfield  {journal} {\bibinfo  {journal} {Phys. Rev. B}\ }\textbf {\bibinfo
  {volume} {66}},\ \bibinfo {pages} {024407} (\bibinfo {year}
  {2002})}\BibitemShut {NoStop}%
\bibitem [{\citenamefont {Xu}\ \emph {et~al.}(1996)\citenamefont {Xu},
  \citenamefont {DiTusa}, \citenamefont {Ito}, \citenamefont {Oka},
  \citenamefont {Takagi}, \citenamefont {Broholm},\ and\ \citenamefont
  {Aeppli}}]{Xu96:54}%
  \BibitemOpen
  \bibfield  {author} {\bibinfo {author} {\bibfnamefont {G.}~\bibnamefont
  {Xu}}, \bibinfo {author} {\bibfnamefont {J.~F.}\ \bibnamefont {DiTusa}},
  \bibinfo {author} {\bibfnamefont {T.}~\bibnamefont {Ito}}, \bibinfo {author}
  {\bibfnamefont {K.}~\bibnamefont {Oka}}, \bibinfo {author} {\bibfnamefont
  {H.}~\bibnamefont {Takagi}}, \bibinfo {author} {\bibfnamefont
  {C.}~\bibnamefont {Broholm}}, \ and\ \bibinfo {author} {\bibfnamefont
  {G.}~\bibnamefont {Aeppli}},\ }\href@noop {} {\bibfield  {journal} {\bibinfo
  {journal} {Phys. Rev. B}\ }\textbf {\bibinfo {volume} {54}},\ \bibinfo
  {pages} {R6827(R)} (\bibinfo {year} {1996})}\BibitemShut {NoStop}%
\bibitem [{\citenamefont {Balents}(2010)}]{Balents10;464}%
  \BibitemOpen
  \bibfield  {author} {\bibinfo {author} {\bibfnamefont {L.}~\bibnamefont
  {Balents}},\ }\href@noop {} {\bibfield  {journal} {\bibinfo  {journal}
  {Nature}\ }\textbf {\bibinfo {volume} {464}},\ \bibinfo {pages} {199}
  (\bibinfo {year} {2010})}\BibitemShut {NoStop}%
\bibitem [{\citenamefont {Nakatsuji}\ \emph {et~al.}(2005)\citenamefont
  {Nakatsuji}, \citenamefont {Nambu}, \citenamefont {Tonomura}, \citenamefont
  {Sakai}, \citenamefont {Jonas}, \citenamefont {Broholm}, \citenamefont
  {Tsunetsugu}, \citenamefont {Qiu},\ and\ \citenamefont
  {Maeno}}]{Nakatsuji05:309}%
  \BibitemOpen
  \bibfield  {author} {\bibinfo {author} {\bibfnamefont {S.}~\bibnamefont
  {Nakatsuji}}, \bibinfo {author} {\bibfnamefont {Y.}~\bibnamefont {Nambu}},
  \bibinfo {author} {\bibfnamefont {H.}~\bibnamefont {Tonomura}}, \bibinfo
  {author} {\bibfnamefont {O.}~\bibnamefont {Sakai}}, \bibinfo {author}
  {\bibfnamefont {S.}~\bibnamefont {Jonas}}, \bibinfo {author} {\bibfnamefont
  {C.}~\bibnamefont {Broholm}}, \bibinfo {author} {\bibfnamefont
  {H.}~\bibnamefont {Tsunetsugu}}, \bibinfo {author} {\bibfnamefont
  {Y.}~\bibnamefont {Qiu}}, \ and\ \bibinfo {author} {\bibfnamefont
  {Y.}~\bibnamefont {Maeno}},\ }\href@noop {} {\bibfield  {journal} {\bibinfo
  {journal} {Science}\ }\textbf {\bibinfo {volume} {309}},\ \bibinfo {pages}
  {1697} (\bibinfo {year} {2005})}\BibitemShut {NoStop}%
\bibitem [{\citenamefont {Nambu}\ \emph {et~al.}(2006)\citenamefont {Nambu},
  \citenamefont {Nakatsuji},\ and\ \citenamefont {Maeno}}]{Nambu06:75}%
  \BibitemOpen
  \bibfield  {author} {\bibinfo {author} {\bibfnamefont {Y.}~\bibnamefont
  {Nambu}}, \bibinfo {author} {\bibfnamefont {S.}~\bibnamefont {Nakatsuji}}, \
  and\ \bibinfo {author} {\bibfnamefont {Y.}~\bibnamefont {Maeno}},\
  }\href@noop {} {\bibfield  {journal} {\bibinfo  {journal} {J. Phys. Soc.
  Jpn.}\ }\textbf {\bibinfo {volume} {75}},\ \bibinfo {pages} {043711}
  (\bibinfo {year} {2006})}\BibitemShut {NoStop}%
\bibitem [{\citenamefont {Stock}\ \emph {et~al.}(2010)\citenamefont {Stock},
  \citenamefont {Jonas}, \citenamefont {Broholm}, \citenamefont {Nakatsuji},
  \citenamefont {Nambu}, \citenamefont {Onuma}, \citenamefont {Maeno},\ and\
  \citenamefont {Chung}}]{Stock10:105}%
  \BibitemOpen
  \bibfield  {author} {\bibinfo {author} {\bibfnamefont {C.}~\bibnamefont
  {Stock}}, \bibinfo {author} {\bibfnamefont {S.}~\bibnamefont {Jonas}},
  \bibinfo {author} {\bibfnamefont {C.}~\bibnamefont {Broholm}}, \bibinfo
  {author} {\bibfnamefont {S.}~\bibnamefont {Nakatsuji}}, \bibinfo {author}
  {\bibfnamefont {Y.}~\bibnamefont {Nambu}}, \bibinfo {author} {\bibfnamefont
  {K.}~\bibnamefont {Onuma}}, \bibinfo {author} {\bibfnamefont
  {Y.}~\bibnamefont {Maeno}}, \ and\ \bibinfo {author} {\bibfnamefont {J.~H.}\
  \bibnamefont {Chung}},\ }\href@noop {} {\bibfield  {journal} {\bibinfo
  {journal} {Phys. Rev. Lett.}\ }\textbf {\bibinfo {volume} {105}},\ \bibinfo
  {pages} {037402} (\bibinfo {year} {2010})}\BibitemShut {NoStop}%
\bibitem [{\citenamefont {Bhattacharjee}\ \emph {et~al.}(2006)\citenamefont
  {Bhattacharjee}, \citenamefont {Shenoy},\ and\ \citenamefont
  {Senthil}}]{Bhattacharjee06;74}%
  \BibitemOpen
  \bibfield  {author} {\bibinfo {author} {\bibfnamefont {S.}~\bibnamefont
  {Bhattacharjee}}, \bibinfo {author} {\bibfnamefont {V.~B.}\ \bibnamefont
  {Shenoy}}, \ and\ \bibinfo {author} {\bibfnamefont {T.}~\bibnamefont
  {Senthil}},\ }\href@noop {} {\bibfield  {journal} {\bibinfo  {journal} {Phys.
  Rev. B}\ }\textbf {\bibinfo {volume} {74}},\ \bibinfo {pages} {092406}
  (\bibinfo {year} {2006})}\BibitemShut {NoStop}%
\bibitem [{\citenamefont {Eerenstein}\ \emph {et~al.}(2006)\citenamefont
  {Eerenstein}, \citenamefont {Mathur},\ and\ \citenamefont
  {Scott}}]{Eerenstein06:442}%
  \BibitemOpen
  \bibfield  {author} {\bibinfo {author} {\bibfnamefont {W.}~\bibnamefont
  {Eerenstein}}, \bibinfo {author} {\bibfnamefont {N.~D.}\ \bibnamefont
  {Mathur}}, \ and\ \bibinfo {author} {\bibfnamefont {J.~F.}\ \bibnamefont
  {Scott}},\ }\href@noop {} {\bibfield  {journal} {\bibinfo  {journal}
  {Nature}\ }\textbf {\bibinfo {volume} {442}},\ \bibinfo {pages} {759}
  (\bibinfo {year} {2006})}\BibitemShut {NoStop}%
\bibitem [{\citenamefont {Cheong}\ and\ \citenamefont
  {Mostovoy}(2007)}]{Cheong07:6}%
  \BibitemOpen
  \bibfield  {author} {\bibinfo {author} {\bibfnamefont {S.-W.}\ \bibnamefont
  {Cheong}}\ and\ \bibinfo {author} {\bibfnamefont {M.}~\bibnamefont
  {Mostovoy}},\ }\href@noop {} {\bibfield  {journal} {\bibinfo  {journal}
  {Nature Materials}\ }\textbf {\bibinfo {volume} {6}},\ \bibinfo {pages} {13}
  (\bibinfo {year} {2007})}\BibitemShut {NoStop}%
\bibitem [{\citenamefont {Oshikawa}\ \emph {et~al.}(1997)\citenamefont
  {Oshikawa}, \citenamefont {Yamanaka},\ and\ \citenamefont
  {Affleck}}]{Oshikawa97:78}%
  \BibitemOpen
  \bibfield  {author} {\bibinfo {author} {\bibfnamefont {M.}~\bibnamefont
  {Oshikawa}}, \bibinfo {author} {\bibfnamefont {M.}~\bibnamefont {Yamanaka}},
  \ and\ \bibinfo {author} {\bibfnamefont {I.}~\bibnamefont {Affleck}},\
  }\href@noop {} {\bibfield  {journal} {\bibinfo  {journal} {Phys. Rev. Lett.}\
  }\textbf {\bibinfo {volume} {78}},\ \bibinfo {pages} {1984} (\bibinfo {year}
  {1997})}\BibitemShut {NoStop}%
\bibitem [{\citenamefont {Kikuchi}\ \emph {et~al.}(2005)\citenamefont
  {Kikuchi}, \citenamefont {Fujii}, \citenamefont {Chiba}, \citenamefont
  {Mitsudo}, \citenamefont {Idehara}, \citenamefont {Tonegawa}, \citenamefont
  {Okamoto}, \citenamefont {Sakai}, \citenamefont {Kuwai},\ and\ \citenamefont
  {Ohta}}]{Kikuchi05:94}%
  \BibitemOpen
  \bibfield  {author} {\bibinfo {author} {\bibfnamefont {H.}~\bibnamefont
  {Kikuchi}}, \bibinfo {author} {\bibfnamefont {Y.}~\bibnamefont {Fujii}},
  \bibinfo {author} {\bibfnamefont {M.}~\bibnamefont {Chiba}}, \bibinfo
  {author} {\bibfnamefont {S.}~\bibnamefont {Mitsudo}}, \bibinfo {author}
  {\bibfnamefont {T.}~\bibnamefont {Idehara}}, \bibinfo {author} {\bibfnamefont
  {T.}~\bibnamefont {Tonegawa}}, \bibinfo {author} {\bibfnamefont
  {K.}~\bibnamefont {Okamoto}}, \bibinfo {author} {\bibfnamefont
  {T.}~\bibnamefont {Sakai}}, \bibinfo {author} {\bibfnamefont
  {T.}~\bibnamefont {Kuwai}}, \ and\ \bibinfo {author} {\bibfnamefont
  {H.}~\bibnamefont {Ohta}},\ }\href@noop {} {\bibfield  {journal} {\bibinfo
  {journal} {Phys. Rev. Lett.}\ }\textbf {\bibinfo {volume} {94}},\ \bibinfo
  {pages} {227201} (\bibinfo {year} {2005})}\BibitemShut {NoStop}%
\bibitem [{\citenamefont {Kageyama}\ \emph
  {et~al.}(1997{\natexlab{a}})\citenamefont {Kageyama}, \citenamefont
  {Yoshimura}, \citenamefont {Kosuge}, \citenamefont {Azuma}, \citenamefont
  {Takano}, \citenamefont {Mitamura},\ and\ \citenamefont
  {Goto}}]{Kageyama97:66}%
  \BibitemOpen
  \bibfield  {author} {\bibinfo {author} {\bibfnamefont {H.}~\bibnamefont
  {Kageyama}}, \bibinfo {author} {\bibfnamefont {K.}~\bibnamefont {Yoshimura}},
  \bibinfo {author} {\bibfnamefont {K.}~\bibnamefont {Kosuge}}, \bibinfo
  {author} {\bibfnamefont {M.}~\bibnamefont {Azuma}}, \bibinfo {author}
  {\bibfnamefont {M.}~\bibnamefont {Takano}}, \bibinfo {author} {\bibfnamefont
  {H.}~\bibnamefont {Mitamura}}, \ and\ \bibinfo {author} {\bibfnamefont
  {T.}~\bibnamefont {Goto}},\ }\href@noop {} {\bibfield  {journal} {\bibinfo
  {journal} {J. Phys. Soc. Jpn.}\ }\textbf {\bibinfo {volume} {66}},\ \bibinfo
  {pages} {3996} (\bibinfo {year} {1997}{\natexlab{a}})}\BibitemShut {NoStop}%
\bibitem [{\citenamefont {Kageyama}\ \emph
  {et~al.}(1997{\natexlab{b}})\citenamefont {Kageyama}, \citenamefont
  {Yoshimura}, \citenamefont {Kosuge}, \citenamefont {Azuma}, \citenamefont
  {Takano}, \citenamefont {Mitamura},\ and\ \citenamefont
  {Goto}}]{Kageyama97:66_2}%
  \BibitemOpen
  \bibfield  {author} {\bibinfo {author} {\bibfnamefont {H.}~\bibnamefont
  {Kageyama}}, \bibinfo {author} {\bibfnamefont {K.}~\bibnamefont {Yoshimura}},
  \bibinfo {author} {\bibfnamefont {K.}~\bibnamefont {Kosuge}}, \bibinfo
  {author} {\bibfnamefont {M.}~\bibnamefont {Azuma}}, \bibinfo {author}
  {\bibfnamefont {M.}~\bibnamefont {Takano}}, \bibinfo {author} {\bibfnamefont
  {H.}~\bibnamefont {Mitamura}}, \ and\ \bibinfo {author} {\bibfnamefont
  {T.}~\bibnamefont {Goto}},\ }\href@noop {} {\bibfield  {journal} {\bibinfo
  {journal} {J. Phys. Soc. Jpn.}\ }\textbf {\bibinfo {volume} {66}},\ \bibinfo
  {pages} {1607} (\bibinfo {year} {1997}{\natexlab{b}})}\BibitemShut {NoStop}%
\bibitem [{\citenamefont {Hardy}\ \emph {et~al.}(2004)\citenamefont {Hardy},
  \citenamefont {Lees}, \citenamefont {Petrenko}, \citenamefont {McK.Paul},
  \citenamefont {Flahaut}, \citenamefont {Hebert},\ and\ \citenamefont
  {Maignan}}]{Hardy04:70}%
  \BibitemOpen
  \bibfield  {author} {\bibinfo {author} {\bibfnamefont {V.}~\bibnamefont
  {Hardy}}, \bibinfo {author} {\bibfnamefont {M.~R.}\ \bibnamefont {Lees}},
  \bibinfo {author} {\bibfnamefont {O.~A.}\ \bibnamefont {Petrenko}}, \bibinfo
  {author} {\bibfnamefont {D.}~\bibnamefont {McK.Paul}}, \bibinfo {author}
  {\bibfnamefont {D.}~\bibnamefont {Flahaut}}, \bibinfo {author} {\bibfnamefont
  {S.}~\bibnamefont {Hebert}}, \ and\ \bibinfo {author} {\bibfnamefont
  {A.}~\bibnamefont {Maignan}},\ }\href@noop {} {\bibfield  {journal} {\bibinfo
   {journal} {Phys. Rev. B}\ }\textbf {\bibinfo {volume} {70}},\ \bibinfo
  {pages} {064424} (\bibinfo {year} {2004})}\BibitemShut {NoStop}%
\bibitem [{\citenamefont {Niitaka}\ \emph {et~al.}(2001)\citenamefont
  {Niitaka}, \citenamefont {Yoshimura}, \citenamefont {Kosuge}, \citenamefont
  {Nishi},\ and\ \citenamefont {Kakurai}}]{Niitaka01:87}%
  \BibitemOpen
  \bibfield  {author} {\bibinfo {author} {\bibfnamefont {S.}~\bibnamefont
  {Niitaka}}, \bibinfo {author} {\bibfnamefont {K.}~\bibnamefont {Yoshimura}},
  \bibinfo {author} {\bibfnamefont {K.}~\bibnamefont {Kosuge}}, \bibinfo
  {author} {\bibfnamefont {M.}~\bibnamefont {Nishi}}, \ and\ \bibinfo {author}
  {\bibfnamefont {K.}~\bibnamefont {Kakurai}},\ }\href@noop {} {\bibfield
  {journal} {\bibinfo  {journal} {Phys. Rev. Lett.}\ }\textbf {\bibinfo
  {volume} {87}},\ \bibinfo {pages} {177202} (\bibinfo {year}
  {2001})}\BibitemShut {NoStop}%
\bibitem [{\citenamefont {Sampathkumaran}\ and\ \citenamefont
  {Niazi}(2002)}]{Samo02:65}%
  \BibitemOpen
  \bibfield  {author} {\bibinfo {author} {\bibfnamefont {E.~V.}\ \bibnamefont
  {Sampathkumaran}}\ and\ \bibinfo {author} {\bibfnamefont {A.}~\bibnamefont
  {Niazi}},\ }\href@noop {} {\bibfield  {journal} {\bibinfo  {journal} {Phys.
  Rev. B}\ }\textbf {\bibinfo {volume} {65}},\ \bibinfo {pages} {180401}
  (\bibinfo {year} {2002})}\BibitemShut {NoStop}%
\bibitem [{\citenamefont {Hardy}\ \emph {et~al.}(2003)\citenamefont {Hardy},
  \citenamefont {Lees}, \citenamefont {Maignan}, \citenamefont {Hebert},
  \citenamefont {Flahaut}, \citenamefont {Martin},\ and\ \citenamefont
  {Paul}}]{Hardy03:15}%
  \BibitemOpen
  \bibfield  {author} {\bibinfo {author} {\bibfnamefont {V.}~\bibnamefont
  {Hardy}}, \bibinfo {author} {\bibfnamefont {M.~R.}\ \bibnamefont {Lees}},
  \bibinfo {author} {\bibfnamefont {A.}~\bibnamefont {Maignan}}, \bibinfo
  {author} {\bibfnamefont {S.}~\bibnamefont {Hebert}}, \bibinfo {author}
  {\bibfnamefont {D.}~\bibnamefont {Flahaut}}, \bibinfo {author} {\bibfnamefont
  {C.}~\bibnamefont {Martin}}, \ and\ \bibinfo {author} {\bibfnamefont {D.~M.}\
  \bibnamefont {Paul}},\ }\href@noop {} {\bibfield  {journal} {\bibinfo
  {journal} {J. Phys. Condens. Matter}\ }\textbf {\bibinfo {volume} {15}},\
  \bibinfo {pages} {5737} (\bibinfo {year} {2003})}\BibitemShut {NoStop}%
\bibitem [{\citenamefont {Hardy}\ \emph {et~al.}(2006)\citenamefont {Hardy},
  \citenamefont {Martin}, \citenamefont {Martinet},\ and\ \citenamefont
  {Andre}}]{Hardy06:74}%
  \BibitemOpen
  \bibfield  {author} {\bibinfo {author} {\bibfnamefont {V.}~\bibnamefont
  {Hardy}}, \bibinfo {author} {\bibfnamefont {C.}~\bibnamefont {Martin}},
  \bibinfo {author} {\bibfnamefont {G.}~\bibnamefont {Martinet}}, \ and\
  \bibinfo {author} {\bibfnamefont {G.}~\bibnamefont {Andre}},\ }\href@noop {}
  {\bibfield  {journal} {\bibinfo  {journal} {Phys. Rev. B}\ }\textbf {\bibinfo
  {volume} {74}},\ \bibinfo {pages} {064413} (\bibinfo {year}
  {2006})}\BibitemShut {NoStop}%
\bibitem [{\citenamefont {Maignan}\ \emph {et~al.}(2004)\citenamefont
  {Maignan}, \citenamefont {Hardy}, \citenamefont {Hebert}, \citenamefont
  {Drillon}, \citenamefont {Lees}, \citenamefont {Petrenko}, \citenamefont
  {Paul},\ and\ \citenamefont {Khomskii}}]{Maignan04:14}%
  \BibitemOpen
  \bibfield  {author} {\bibinfo {author} {\bibfnamefont {A.}~\bibnamefont
  {Maignan}}, \bibinfo {author} {\bibfnamefont {V.}~\bibnamefont {Hardy}},
  \bibinfo {author} {\bibfnamefont {S.}~\bibnamefont {Hebert}}, \bibinfo
  {author} {\bibfnamefont {M.}~\bibnamefont {Drillon}}, \bibinfo {author}
  {\bibfnamefont {M.~R.}\ \bibnamefont {Lees}}, \bibinfo {author}
  {\bibfnamefont {O.}~\bibnamefont {Petrenko}}, \bibinfo {author}
  {\bibfnamefont {D.~M.}\ \bibnamefont {Paul}}, \ and\ \bibinfo {author}
  {\bibfnamefont {D.}~\bibnamefont {Khomskii}},\ }\href@noop {} {\bibfield
  {journal} {\bibinfo  {journal} {J. Mater. Chem.}\ }\textbf {\bibinfo {volume}
  {14}},\ \bibinfo {pages} {1231} (\bibinfo {year} {2004})}\BibitemShut
  {NoStop}%
\bibitem [{\citenamefont {Kudasov}(2006)}]{Kudasov06:96}%
  \BibitemOpen
  \bibfield  {author} {\bibinfo {author} {\bibfnamefont {Y.~B.}\ \bibnamefont
  {Kudasov}},\ }\href@noop {} {\bibfield  {journal} {\bibinfo  {journal} {Phys.
  Rev. Lett.}\ }\textbf {\bibinfo {volume} {96}},\ \bibinfo {pages} {027212}
  (\bibinfo {year} {2006})}\BibitemShut {NoStop}%
\bibitem [{\citenamefont {Fishman}(2011)}]{Fishman11:106}%
  \BibitemOpen
  \bibfield  {author} {\bibinfo {author} {\bibfnamefont {R.~S.}\ \bibnamefont
  {Fishman}},\ }\href@noop {} {\bibfield  {journal} {\bibinfo  {journal} {Phys.
  Rev. Lett.}\ }\textbf {\bibinfo {volume} {106}},\ \bibinfo {pages} {037206}
  (\bibinfo {year} {2011})}\BibitemShut {NoStop}%
\bibitem [{\citenamefont {Markkula}\ \emph
  {et~al.}(2012{\natexlab{a}})\citenamefont {Markkula}, \citenamefont
  {Arevalo-Lopez},\ and\ \citenamefont {Attfield}}]{Markkula12:86}%
  \BibitemOpen
  \bibfield  {author} {\bibinfo {author} {\bibfnamefont {M.}~\bibnamefont
  {Markkula}}, \bibinfo {author} {\bibfnamefont {A.~M.}\ \bibnamefont
  {Arevalo-Lopez}}, \ and\ \bibinfo {author} {\bibfnamefont {J.~P.}\
  \bibnamefont {Attfield}},\ }\href@noop {} {\bibfield  {journal} {\bibinfo
  {journal} {Phys. Rev. B}\ }\textbf {\bibinfo {volume} {86}},\ \bibinfo
  {pages} {134401} (\bibinfo {year} {2012}{\natexlab{a}})}\BibitemShut
  {NoStop}%
\bibitem [{\citenamefont {Okamoto}\ \emph {et~al.}(2003)\citenamefont
  {Okamoto}, \citenamefont {Tonegawa},\ and\ \citenamefont
  {Kaburagi}}]{Okamoto03:15}%
  \BibitemOpen
  \bibfield  {author} {\bibinfo {author} {\bibfnamefont {K.}~\bibnamefont
  {Okamoto}}, \bibinfo {author} {\bibfnamefont {T.}~\bibnamefont {Tonegawa}}, \
  and\ \bibinfo {author} {\bibfnamefont {M.}~\bibnamefont {Kaburagi}},\
  }\href@noop {} {\bibfield  {journal} {\bibinfo  {journal} {J. Phys.: Condens.
  Matter}\ }\textbf {\bibinfo {volume} {15}},\ \bibinfo {pages} {5979}
  (\bibinfo {year} {2003})}\BibitemShut {NoStop}%
\bibitem [{\citenamefont {Rule}\ \emph {et~al.}(2008)\citenamefont {Rule},
  \citenamefont {Wolter}, \citenamefont {Sullow}, \citenamefont {Tennant},
  \citenamefont {Bruhl}, \citenamefont {Kohler}, \citenamefont {Wolf},
  \citenamefont {Lang},\ and\ \citenamefont {Schreuer}}]{Rule08:100}%
  \BibitemOpen
  \bibfield  {author} {\bibinfo {author} {\bibfnamefont {K.~C.}\ \bibnamefont
  {Rule}}, \bibinfo {author} {\bibfnamefont {A.~U.~B.}\ \bibnamefont {Wolter}},
  \bibinfo {author} {\bibfnamefont {S.}~\bibnamefont {Sullow}}, \bibinfo
  {author} {\bibfnamefont {D.~A.}\ \bibnamefont {Tennant}}, \bibinfo {author}
  {\bibfnamefont {A.}~\bibnamefont {Bruhl}}, \bibinfo {author} {\bibfnamefont
  {S.}~\bibnamefont {Kohler}}, \bibinfo {author} {\bibfnamefont
  {B.}~\bibnamefont {Wolf}}, \bibinfo {author} {\bibfnamefont {M.}~\bibnamefont
  {Lang}}, \ and\ \bibinfo {author} {\bibfnamefont {J.}~\bibnamefont
  {Schreuer}},\ }\href@noop {} {\bibfield  {journal} {\bibinfo  {journal}
  {Phys. Rev. Lett.}\ }\textbf {\bibinfo {volume} {100}},\ \bibinfo {pages}
  {117202} (\bibinfo {year} {2008})}\BibitemShut {NoStop}%
\bibitem [{\citenamefont {Lenertz}\ \emph {et~al.}(2011)\citenamefont
  {Lenertz}, \citenamefont {Alaria}, \citenamefont {Stoeffler}, \citenamefont
  {Colis},\ and\ \citenamefont {Dinia}}]{Lenertz11:115}%
  \BibitemOpen
  \bibfield  {author} {\bibinfo {author} {\bibfnamefont {M.}~\bibnamefont
  {Lenertz}}, \bibinfo {author} {\bibfnamefont {J.}~\bibnamefont {Alaria}},
  \bibinfo {author} {\bibfnamefont {D.}~\bibnamefont {Stoeffler}}, \bibinfo
  {author} {\bibfnamefont {S.}~\bibnamefont {Colis}}, \ and\ \bibinfo {author}
  {\bibfnamefont {A.}~\bibnamefont {Dinia}},\ }\href@noop {} {\bibfield
  {journal} {\bibinfo  {journal} {J. Phys. Chem. C}\ }\textbf {\bibinfo
  {volume} {115}},\ \bibinfo {pages} {17190} (\bibinfo {year}
  {2011})}\BibitemShut {NoStop}%
\bibitem [{\citenamefont {Saul}\ \emph {et~al.}(2013)\citenamefont {Saul},
  \citenamefont {Vodenicarevic},\ and\ \citenamefont {Radtke}}]{Saul13:87}%
  \BibitemOpen
  \bibfield  {author} {\bibinfo {author} {\bibfnamefont {A.}~\bibnamefont
  {Saul}}, \bibinfo {author} {\bibfnamefont {D.}~\bibnamefont {Vodenicarevic}},
  \ and\ \bibinfo {author} {\bibfnamefont {G.}~\bibnamefont {Radtke}},\
  }\href@noop {} {\bibfield  {journal} {\bibinfo  {journal} {Phys. Rev. B}\
  }\textbf {\bibinfo {volume} {87}},\ \bibinfo {pages} {024403} (\bibinfo
  {year} {2013})}\BibitemShut {NoStop}%
\bibitem [{\citenamefont {Hollmann}\ \emph {et~al.}(2014)\citenamefont
  {Hollmann}, \citenamefont {Agrestini}, \citenamefont {Hu}, \citenamefont
  {He}, \citenamefont {Schmidt}, \citenamefont {Kuo}, \citenamefont {Rotter},
  \citenamefont {Nugroho}, \citenamefont {Sessi}, \citenamefont {Tanaka},
  \citenamefont {Brookes},\ and\ \citenamefont {Tjeng}}]{Hollmann14:89}%
  \BibitemOpen
  \bibfield  {author} {\bibinfo {author} {\bibfnamefont {N.}~\bibnamefont
  {Hollmann}}, \bibinfo {author} {\bibfnamefont {S.}~\bibnamefont {Agrestini}},
  \bibinfo {author} {\bibfnamefont {Z.}~\bibnamefont {Hu}}, \bibinfo {author}
  {\bibfnamefont {Z.}~\bibnamefont {He}}, \bibinfo {author} {\bibfnamefont
  {M.}~\bibnamefont {Schmidt}}, \bibinfo {author} {\bibfnamefont {C.~Y.}\
  \bibnamefont {Kuo}}, \bibinfo {author} {\bibfnamefont {M.}~\bibnamefont
  {Rotter}}, \bibinfo {author} {\bibfnamefont {A.~A.}\ \bibnamefont {Nugroho}},
  \bibinfo {author} {\bibfnamefont {V.}~\bibnamefont {Sessi}}, \bibinfo
  {author} {\bibfnamefont {A.}~\bibnamefont {Tanaka}}, \bibinfo {author}
  {\bibfnamefont {N.~B.}\ \bibnamefont {Brookes}}, \ and\ \bibinfo {author}
  {\bibfnamefont {L.~H.}\ \bibnamefont {Tjeng}},\ }\href@noop {} {\bibfield
  {journal} {\bibinfo  {journal} {Phys. Rev. B}\ }\textbf {\bibinfo {volume}
  {89}},\ \bibinfo {pages} {201101(R)} (\bibinfo {year} {2014})}\BibitemShut
  {NoStop}%
\bibitem [{\citenamefont {Markkula}\ \emph
  {et~al.}(2012{\natexlab{b}})\citenamefont {Markkula}, \citenamefont
  {Arevalo-Lopez},\ and\ \citenamefont {Attfield}}]{Marrkkula12:192}%
  \BibitemOpen
  \bibfield  {author} {\bibinfo {author} {\bibfnamefont {M.}~\bibnamefont
  {Markkula}}, \bibinfo {author} {\bibfnamefont {A.~M.}\ \bibnamefont
  {Arevalo-Lopez}}, \ and\ \bibinfo {author} {\bibfnamefont {J.~P.}\
  \bibnamefont {Attfield}},\ }\href@noop {} {\bibfield  {journal} {\bibinfo
  {journal} {J. Sol. State Chem.}\ }\textbf {\bibinfo {volume} {192}},\
  \bibinfo {pages} {390} (\bibinfo {year} {2012}{\natexlab{b}})}\BibitemShut
  {NoStop}%
\bibitem [{\citenamefont {Singh}\ \emph {et~al.}(2012)\citenamefont {Singh},
  \citenamefont {Maignan}, \citenamefont {Pelloquin},\ and\ \citenamefont
  {Simon}}]{Singh12:22}%
  \BibitemOpen
  \bibfield  {author} {\bibinfo {author} {\bibfnamefont {J.}~\bibnamefont
  {Singh}}, \bibinfo {author} {\bibfnamefont {A.}~\bibnamefont {Maignan}},
  \bibinfo {author} {\bibfnamefont {D.}~\bibnamefont {Pelloquin}}, \ and\
  \bibinfo {author} {\bibfnamefont {C.}~\bibnamefont {Simon}},\ }\href@noop {}
  {\bibfield  {journal} {\bibinfo  {journal} {J. Mat. Chem.}\ }\textbf
  {\bibinfo {volume} {22}},\ \bibinfo {pages} {6436} (\bibinfo {year}
  {2012})}\BibitemShut {NoStop}%
\bibitem [{\citenamefont {Kim}\ \emph {et~al.}(2012)\citenamefont {Kim},
  \citenamefont {Kim}, \citenamefont {Kim}, \citenamefont {Choi}, \citenamefont
  {Park}, \citenamefont {Jeong},\ and\ \citenamefont {Min}}]{Kim12:85}%
  \BibitemOpen
  \bibfield  {author} {\bibinfo {author} {\bibfnamefont {B.}~\bibnamefont
  {Kim}}, \bibinfo {author} {\bibfnamefont {B.~H.}\ \bibnamefont {Kim}},
  \bibinfo {author} {\bibfnamefont {K.}~\bibnamefont {Kim}}, \bibinfo {author}
  {\bibfnamefont {H.~C.}\ \bibnamefont {Choi}}, \bibinfo {author}
  {\bibfnamefont {S.~Y.}\ \bibnamefont {Park}}, \bibinfo {author}
  {\bibfnamefont {Y.~H.}\ \bibnamefont {Jeong}}, \ and\ \bibinfo {author}
  {\bibfnamefont {B.~I.}\ \bibnamefont {Min}},\ }\href@noop {} {\bibfield
  {journal} {\bibinfo  {journal} {Phys. Rev. B}\ }\textbf {\bibinfo {volume}
  {85}},\ \bibinfo {pages} {220407(R)} (\bibinfo {year} {2012})}\BibitemShut
  {NoStop}%
\bibitem [{\citenamefont {Cowley}\ \emph {et~al.}(2013)\citenamefont {Cowley},
  \citenamefont {Buyers}, \citenamefont {Stock}, \citenamefont {Yamani},
  \citenamefont {Frost}, \citenamefont {Taylor},\ and\ \citenamefont
  {Prabhakaran}}]{Cowley13:88}%
  \BibitemOpen
  \bibfield  {author} {\bibinfo {author} {\bibfnamefont {R.~A.}\ \bibnamefont
  {Cowley}}, \bibinfo {author} {\bibfnamefont {W.~J.~L.}\ \bibnamefont
  {Buyers}}, \bibinfo {author} {\bibfnamefont {C.}~\bibnamefont {Stock}},
  \bibinfo {author} {\bibfnamefont {Z.}~\bibnamefont {Yamani}}, \bibinfo
  {author} {\bibfnamefont {C.}~\bibnamefont {Frost}}, \bibinfo {author}
  {\bibfnamefont {J.~W.}\ \bibnamefont {Taylor}}, \ and\ \bibinfo {author}
  {\bibfnamefont {D.}~\bibnamefont {Prabhakaran}},\ }\href@noop {} {\bibfield
  {journal} {\bibinfo  {journal} {Phys. Rev. B}\ }\textbf {\bibinfo {volume}
  {88}},\ \bibinfo {pages} {205117} (\bibinfo {year} {2013})}\BibitemShut
  {NoStop}%
\bibitem [{\citenamefont {Cowley}\ \emph {et~al.}(1973)\citenamefont {Cowley},
  \citenamefont {Buyers}, \citenamefont {Martel},\ and\ \citenamefont
  {Stevensons}}]{Cowley73:6}%
  \BibitemOpen
  \bibfield  {author} {\bibinfo {author} {\bibfnamefont {R.~A.}\ \bibnamefont
  {Cowley}}, \bibinfo {author} {\bibfnamefont {W.~J.~L.}\ \bibnamefont
  {Buyers}}, \bibinfo {author} {\bibfnamefont {P.}~\bibnamefont {Martel}}, \
  and\ \bibinfo {author} {\bibfnamefont {R.~W.~H.}\ \bibnamefont
  {Stevensons}},\ }\href@noop {} {\bibfield  {journal} {\bibinfo  {journal} {J.
  Phys. C: Solid State Phys.}\ }\textbf {\bibinfo {volume} {6}},\ \bibinfo
  {pages} {2997} (\bibinfo {year} {1973})}\BibitemShut {NoStop}%
\bibitem [{\citenamefont {Kant}\ \emph {et~al.}(2008)\citenamefont {Kant},
  \citenamefont {Rudolf}, \citenamefont {Schrettle}, \citenamefont {Mayr},
  \citenamefont {Deisenhofer}, \citenamefont {Lunkenheimer}, \citenamefont
  {Eremin},\ and\ \citenamefont {Loidl}}]{Kant08:78}%
  \BibitemOpen
  \bibfield  {author} {\bibinfo {author} {\bibfnamefont {C.}~\bibnamefont
  {Kant}}, \bibinfo {author} {\bibfnamefont {T.}~\bibnamefont {Rudolf}},
  \bibinfo {author} {\bibfnamefont {F.}~\bibnamefont {Schrettle}}, \bibinfo
  {author} {\bibfnamefont {F.}~\bibnamefont {Mayr}}, \bibinfo {author}
  {\bibfnamefont {J.}~\bibnamefont {Deisenhofer}}, \bibinfo {author}
  {\bibfnamefont {P.}~\bibnamefont {Lunkenheimer}}, \bibinfo {author}
  {\bibfnamefont {M.~V.}\ \bibnamefont {Eremin}}, \ and\ \bibinfo {author}
  {\bibfnamefont {A.}~\bibnamefont {Loidl}},\ }\href@noop {} {\bibfield
  {journal} {\bibinfo  {journal} {Phys. Rev. B}\ }\textbf {\bibinfo {volume}
  {78}},\ \bibinfo {pages} {245103} (\bibinfo {year} {2008})}\BibitemShut
  {NoStop}%
\bibitem [{\citenamefont {Khomskii}(2014)}]{Khomskii:book}%
  \BibitemOpen
  \bibfield  {author} {\bibinfo {author} {\bibfnamefont {D.~I.}\ \bibnamefont
  {Khomskii}},\ }\href@noop {} {\emph {\bibinfo {title} {Transition Metal
  Compounds}}}\ (\bibinfo  {publisher} {Cambridge University Press},\ \bibinfo
  {address} {Cambridge},\ \bibinfo {year} {2014})\BibitemShut {NoStop}%
\bibitem [{\citenamefont {Abragam}\ and\ \citenamefont
  {Bleaney}(1986)}]{Abragam:book}%
  \BibitemOpen
  \bibfield  {author} {\bibinfo {author} {\bibfnamefont {A.}~\bibnamefont
  {Abragam}}\ and\ \bibinfo {author} {\bibfnamefont {B.}~\bibnamefont
  {Bleaney}},\ }\href@noop {} {\emph {\bibinfo {title} {Electron paramagnetic
  resonance of transition ions}}}\ (\bibinfo  {publisher} {Dover
  Publications},\ \bibinfo {address} {New York},\ \bibinfo {year}
  {1986})\BibitemShut {NoStop}%
\bibitem [{\citenamefont {Ballhausen}(1962)}]{Ballhausen:book}%
  \BibitemOpen
  \bibfield  {author} {\bibinfo {author} {\bibfnamefont {C.~J.}\ \bibnamefont
  {Ballhausen}},\ }\href@noop {} {\emph {\bibinfo {title} {Ligand Field
  Theory}}}\ (\bibinfo  {publisher} {McGraw-Hill},\ \bibinfo {address} {New
  York},\ \bibinfo {year} {1962})\BibitemShut {NoStop}%
\bibitem [{\citenamefont {McClure}(1959)}]{McClure59:9}%
  \BibitemOpen
  \bibfield  {author} {\bibinfo {author} {\bibfnamefont {D.}~\bibnamefont
  {McClure}},\ }\href@noop {} {\bibfield  {journal} {\bibinfo  {journal} {Sol.
  State. Phys.}\ }\textbf {\bibinfo {volume} {9}},\ \bibinfo {pages} {399}
  (\bibinfo {year} {1959})}\BibitemShut {NoStop}%
\bibitem [{\citenamefont {Hutchings}(1965)}]{Hutchings65:16}%
  \BibitemOpen
  \bibfield  {author} {\bibinfo {author} {\bibfnamefont {M.~T.}\ \bibnamefont
  {Hutchings}},\ }\href@noop {} {\bibfield  {journal} {\bibinfo  {journal}
  {Solid State Phys.}\ }\textbf {\bibinfo {volume} {16}},\ \bibinfo {pages}
  {094434} (\bibinfo {year} {1965})}\BibitemShut {NoStop}%
\bibitem [{\citenamefont {Lines}(1963)}]{Lines63:131}%
  \BibitemOpen
  \bibfield  {author} {\bibinfo {author} {\bibfnamefont {M.~E.}\ \bibnamefont
  {Lines}},\ }\href@noop {} {\bibfield  {journal} {\bibinfo  {journal} {Phys.
  Rev.}\ }\textbf {\bibinfo {volume} {131}},\ \bibinfo {pages} {546} (\bibinfo
  {year} {1963})}\BibitemShut {NoStop}%
\bibitem [{\citenamefont {Kim}\ \emph {et~al.}(2011)\citenamefont {Kim},
  \citenamefont {Sorini}, \citenamefont {Stock}, \citenamefont {Perring},
  \citenamefont {van~den Brink},\ and\ \citenamefont {Devereaux}}]{Kim11:84}%
  \BibitemOpen
  \bibfield  {author} {\bibinfo {author} {\bibfnamefont {Y.-J.}\ \bibnamefont
  {Kim}}, \bibinfo {author} {\bibfnamefont {A.~P.}\ \bibnamefont {Sorini}},
  \bibinfo {author} {\bibfnamefont {C.}~\bibnamefont {Stock}}, \bibinfo
  {author} {\bibfnamefont {T.~G.}\ \bibnamefont {Perring}}, \bibinfo {author}
  {\bibfnamefont {J.}~\bibnamefont {van~den Brink}}, \ and\ \bibinfo {author}
  {\bibfnamefont {T.~P.}\ \bibnamefont {Devereaux}},\ }\href@noop {} {\bibfield
   {journal} {\bibinfo  {journal} {Phys. Rev. B}\ }\textbf {\bibinfo {volume}
  {84}},\ \bibinfo {pages} {085132} (\bibinfo {year} {2011})}\BibitemShut
  {NoStop}%
\bibitem [{\citenamefont {Larson}\ \emph {et~al.}(2007)\citenamefont {Larson},
  \citenamefont {Ku}, \citenamefont {Tischler}, \citenamefont {Lee},
  \citenamefont {Restrepo}, \citenamefont {Eguiluz}, \citenamefont {Zschack},\
  and\ \citenamefont {Finkelstein}}]{Larson07:99}%
  \BibitemOpen
  \bibfield  {author} {\bibinfo {author} {\bibfnamefont {B.~C.}\ \bibnamefont
  {Larson}}, \bibinfo {author} {\bibfnamefont {W.}~\bibnamefont {Ku}}, \bibinfo
  {author} {\bibfnamefont {J.~Z.}\ \bibnamefont {Tischler}}, \bibinfo {author}
  {\bibfnamefont {C.~C.}\ \bibnamefont {Lee}}, \bibinfo {author} {\bibfnamefont
  {O.~D.}\ \bibnamefont {Restrepo}}, \bibinfo {author} {\bibfnamefont {A.~G.}\
  \bibnamefont {Eguiluz}}, \bibinfo {author} {\bibfnamefont {P.}~\bibnamefont
  {Zschack}}, \ and\ \bibinfo {author} {\bibfnamefont {K.~D.}\ \bibnamefont
  {Finkelstein}},\ }\href@noop {} {\bibfield  {journal} {\bibinfo  {journal}
  {Phys. Rev. Lett.}\ }\textbf {\bibinfo {volume} {99}},\ \bibinfo {pages}
  {026401} (\bibinfo {year} {2007})}\BibitemShut {NoStop}%
\bibitem [{\citenamefont {Haverkort}\ \emph {et~al.}(2007)\citenamefont
  {Haverkort}, \citenamefont {Tanaka}, \citenamefont {Tjeng},\ and\
  \citenamefont {Sawatzky}}]{Haverkort07:99}%
  \BibitemOpen
  \bibfield  {author} {\bibinfo {author} {\bibfnamefont {M.~W.}\ \bibnamefont
  {Haverkort}}, \bibinfo {author} {\bibfnamefont {A.}~\bibnamefont {Tanaka}},
  \bibinfo {author} {\bibfnamefont {L.~H.}\ \bibnamefont {Tjeng}}, \ and\
  \bibinfo {author} {\bibfnamefont {G.~A.}\ \bibnamefont {Sawatzky}},\
  }\href@noop {} {\bibfield  {journal} {\bibinfo  {journal} {Phys. Rev. Lett.}\
  }\textbf {\bibinfo {volume} {99}},\ \bibinfo {pages} {257401} (\bibinfo
  {year} {2007})}\BibitemShut {NoStop}%
\bibitem [{\citenamefont {Griffiths}(1964)}]{Griffiths:book}%
  \BibitemOpen
  \bibfield  {author} {\bibinfo {author} {\bibfnamefont {J.}~\bibnamefont
  {Griffiths}},\ }\href@noop {} {\emph {\bibinfo {title} {The Theory of
  Transition-Metal Ions}}}\ (\bibinfo  {publisher} {Cambridge University
  Press},\ \bibinfo {address} {London},\ \bibinfo {year} {1964})\BibitemShut
  {NoStop}%
\bibitem [{\citenamefont {Liehr}(1963)}]{Liehr63:67}%
  \BibitemOpen
  \bibfield  {author} {\bibinfo {author} {\bibfnamefont {A.~D.}\ \bibnamefont
  {Liehr}},\ }\href@noop {} {\bibfield  {journal} {\bibinfo  {journal} {J.
  Phys. Chem.}\ }\textbf {\bibinfo {volume} {67}},\ \bibinfo {pages} {1314}
  (\bibinfo {year} {1963})}\BibitemShut {NoStop}%
\bibitem [{\citenamefont {Decaroli}\ \emph {et~al.}(2015)\citenamefont
  {Decaroli}, \citenamefont {Arevalo-Lopez}, \citenamefont {Woodall},
  \citenamefont {Rodriguez}, \citenamefont {Attfield}, \citenamefont {Parker},\
  and\ \citenamefont {Stock}}]{Decaroli15:71}%
  \BibitemOpen
  \bibfield  {author} {\bibinfo {author} {\bibfnamefont {C.}~\bibnamefont
  {Decaroli}}, \bibinfo {author} {\bibfnamefont {A.~M.}\ \bibnamefont
  {Arevalo-Lopez}}, \bibinfo {author} {\bibfnamefont {C.~H.}\ \bibnamefont
  {Woodall}}, \bibinfo {author} {\bibfnamefont {E.~E.}\ \bibnamefont
  {Rodriguez}}, \bibinfo {author} {\bibfnamefont {J.~P.}\ \bibnamefont
  {Attfield}}, \bibinfo {author} {\bibfnamefont {S.~F.}\ \bibnamefont
  {Parker}}, \ and\ \bibinfo {author} {\bibfnamefont {C.}~\bibnamefont
  {Stock}},\ }\href@noop {} {\bibfield  {journal} {\bibinfo  {journal} {Act.
  Cryst. B}\ }\textbf {\bibinfo {volume} {71}},\ \bibinfo {pages} {20}
  (\bibinfo {year} {2015})}\BibitemShut {NoStop}%
\bibitem [{\citenamefont {Kanamori}(1957{\natexlab{a}})}]{Kanamor_A17:177}%
  \BibitemOpen
  \bibfield  {author} {\bibinfo {author} {\bibfnamefont {J.}~\bibnamefont
  {Kanamori}},\ }\href@noop {} {\bibfield  {journal} {\bibinfo  {journal}
  {Prog. Theor. Phys.}\ }\textbf {\bibinfo {volume} {17}},\ \bibinfo {pages}
  {177} (\bibinfo {year} {1957}{\natexlab{a}})}\BibitemShut {NoStop}%
\bibitem [{\citenamefont {Kanamori}(1957{\natexlab{b}})}]{Kanamor_B17:177}%
  \BibitemOpen
  \bibfield  {author} {\bibinfo {author} {\bibfnamefont {J.}~\bibnamefont
  {Kanamori}},\ }\href@noop {} {\bibfield  {journal} {\bibinfo  {journal}
  {Prog. Theor. Phys.}\ }\textbf {\bibinfo {volume} {17}},\ \bibinfo {pages}
  {197} (\bibinfo {year} {1957}{\natexlab{b}})}\BibitemShut {NoStop}%
\bibitem [{\citenamefont {Walter}(1984)}]{Walter84:45}%
  \BibitemOpen
  \bibfield  {author} {\bibinfo {author} {\bibfnamefont {U.}~\bibnamefont
  {Walter}},\ }\href@noop {} {\bibfield  {journal} {\bibinfo  {journal} {J.
  Phys. Chem. Solids}\ }\textbf {\bibinfo {volume} {45}},\ \bibinfo {pages}
  {401} (\bibinfo {year} {1984})}\BibitemShut {NoStop}%
\bibitem [{\citenamefont {Martel}\ \emph {et~al.}(1968)\citenamefont {Martel},
  \citenamefont {Cowley},\ and\ \citenamefont {Stevenson}}]{Martel68:46}%
  \BibitemOpen
  \bibfield  {author} {\bibinfo {author} {\bibfnamefont {P.}~\bibnamefont
  {Martel}}, \bibinfo {author} {\bibfnamefont {R.~A.}\ \bibnamefont {Cowley}},
  \ and\ \bibinfo {author} {\bibfnamefont {R.~W.~H.}\ \bibnamefont
  {Stevenson}},\ }\href@noop {} {\bibfield  {journal} {\bibinfo  {journal}
  {Can. J. Phys.}\ }\textbf {\bibinfo {volume} {46}},\ \bibinfo {pages} {1355}
  (\bibinfo {year} {1968})}\BibitemShut {NoStop}%
\bibitem [{\citenamefont {Gladney}(1966)}]{Gladney66:146}%
  \BibitemOpen
  \bibfield  {author} {\bibinfo {author} {\bibfnamefont {N.~M.}\ \bibnamefont
  {Gladney}},\ }\href@noop {} {\bibfield  {journal} {\bibinfo  {journal} {Phys.
  Rev.}\ }\textbf {\bibinfo {volume} {146}},\ \bibinfo {pages} {253} (\bibinfo
  {year} {1966})}\BibitemShut {NoStop}%
\bibitem [{\citenamefont {Kimber}\ \emph {et~al.}(2011)\citenamefont {Kimber},
  \citenamefont {Mutka}, \citenamefont {Chatterji}, \citenamefont {Hofmann},
  \citenamefont {Henry}, \citenamefont {Bordallo}, \citenamefont {Argyriou},\
  and\ \citenamefont {Attfield}}]{Kimber11:84}%
  \BibitemOpen
  \bibfield  {author} {\bibinfo {author} {\bibfnamefont {S.~A.~J.}\
  \bibnamefont {Kimber}}, \bibinfo {author} {\bibfnamefont {H.}~\bibnamefont
  {Mutka}}, \bibinfo {author} {\bibfnamefont {T.}~\bibnamefont {Chatterji}},
  \bibinfo {author} {\bibfnamefont {T.}~\bibnamefont {Hofmann}}, \bibinfo
  {author} {\bibfnamefont {P.~F.}\ \bibnamefont {Henry}}, \bibinfo {author}
  {\bibfnamefont {H.~N.}\ \bibnamefont {Bordallo}}, \bibinfo {author}
  {\bibfnamefont {D.~N.}\ \bibnamefont {Argyriou}}, \ and\ \bibinfo {author}
  {\bibfnamefont {J.~P.}\ \bibnamefont {Attfield}},\ }\href@noop {} {\bibfield
  {journal} {\bibinfo  {journal} {Phys. Rev. B}\ }\textbf {\bibinfo {volume}
  {84}},\ \bibinfo {pages} {104425} (\bibinfo {year} {2011})}\BibitemShut
  {NoStop}%
\bibitem [{\citenamefont {Hohenberg}\ and\ \citenamefont
  {Brinkman}(1974)}]{Hohenberg74:10}%
  \BibitemOpen
  \bibfield  {author} {\bibinfo {author} {\bibfnamefont {P.~C.}\ \bibnamefont
  {Hohenberg}}\ and\ \bibinfo {author} {\bibfnamefont {W.~F.}\ \bibnamefont
  {Brinkman}},\ }\href@noop {} {\bibfield  {journal} {\bibinfo  {journal}
  {Phys. Rev. B}\ }\textbf {\bibinfo {volume} {10}},\ \bibinfo {pages} {128}
  (\bibinfo {year} {1974})}\BibitemShut {NoStop}%
\bibitem [{\citenamefont {Zaliznyak}\ and\ \citenamefont
  {Lee}(1996)}]{Zal05:book}%
  \BibitemOpen
  \bibfield  {author} {\bibinfo {author} {\bibfnamefont {I.~A.}\ \bibnamefont
  {Zaliznyak}}\ and\ \bibinfo {author} {\bibfnamefont {S.~H.}\ \bibnamefont
  {Lee}},\ }\href@noop {} {\bibfield  {journal} {\bibinfo  {journal} {Magnetic
  Neutron Scattering in Moder Techniques for Characterizing Magnetic
  Materials}\ }\textbf {\bibinfo {volume} {54}},\ \bibinfo {pages} {R6827(R)}
  (\bibinfo {year} {1996})}\BibitemShut {NoStop}%
\bibitem [{\citenamefont {Stock}\ \emph {et~al.}(2012)\citenamefont {Stock},
  \citenamefont {Rodriguez},\ and\ \citenamefont {Green}}]{Stock12:85}%
  \BibitemOpen
  \bibfield  {author} {\bibinfo {author} {\bibfnamefont {C.}~\bibnamefont
  {Stock}}, \bibinfo {author} {\bibfnamefont {E.~E.}\ \bibnamefont
  {Rodriguez}}, \ and\ \bibinfo {author} {\bibfnamefont {M.~A.}\ \bibnamefont
  {Green}},\ }\href@noop {} {\bibfield  {journal} {\bibinfo  {journal} {Phys.
  Rev. B}\ }\textbf {\bibinfo {volume} {85}},\ \bibinfo {pages} {094507}
  (\bibinfo {year} {2012})}\BibitemShut {NoStop}%
\bibitem [{\citenamefont {Xu}\ \emph {et~al.}(2000)\citenamefont {Xu},
  \citenamefont {Broholm}, \citenamefont {Reich},\ and\ \citenamefont
  {Adams}}]{Xu00:84}%
  \BibitemOpen
  \bibfield  {author} {\bibinfo {author} {\bibfnamefont {G.}~\bibnamefont
  {Xu}}, \bibinfo {author} {\bibfnamefont {C.}~\bibnamefont {Broholm}},
  \bibinfo {author} {\bibfnamefont {D.~H.}\ \bibnamefont {Reich}}, \ and\
  \bibinfo {author} {\bibfnamefont {M.~A.}\ \bibnamefont {Adams}},\ }\href@noop
  {} {\bibfield  {journal} {\bibinfo  {journal} {Phys. Rev. Lett.}\ }\textbf
  {\bibinfo {volume} {84}},\ \bibinfo {pages} {4465} (\bibinfo {year}
  {2000})}\BibitemShut {NoStop}%
\bibitem [{\citenamefont {Stone}\ \emph {et~al.}(2001)\citenamefont {Stone},
  \citenamefont {Zaliznyak}, \citenamefont {Reich},\ and\ \citenamefont
  {Broholm}}]{Stone01:64}%
  \BibitemOpen
  \bibfield  {author} {\bibinfo {author} {\bibfnamefont {M.~B.}\ \bibnamefont
  {Stone}}, \bibinfo {author} {\bibfnamefont {I.}~\bibnamefont {Zaliznyak}},
  \bibinfo {author} {\bibfnamefont {D.~H.}\ \bibnamefont {Reich}}, \ and\
  \bibinfo {author} {\bibfnamefont {C.}~\bibnamefont {Broholm}},\ }\href@noop
  {} {\bibfield  {journal} {\bibinfo  {journal} {Phys. Rev. B}\ }\textbf
  {\bibinfo {volume} {64}},\ \bibinfo {pages} {144405} (\bibinfo {year}
  {2001})}\BibitemShut {NoStop}%
\bibitem [{\citenamefont {Hammar}\ \emph {et~al.}(1998)\citenamefont {Hammar},
  \citenamefont {Reich}, \citenamefont {Broholm},\ and\ \citenamefont
  {Trouw}}]{Hammar98:57}%
  \BibitemOpen
  \bibfield  {author} {\bibinfo {author} {\bibfnamefont {P.~R.}\ \bibnamefont
  {Hammar}}, \bibinfo {author} {\bibfnamefont {D.~H.}\ \bibnamefont {Reich}},
  \bibinfo {author} {\bibfnamefont {C.}~\bibnamefont {Broholm}}, \ and\
  \bibinfo {author} {\bibfnamefont {F.}~\bibnamefont {Trouw}},\ }\href@noop {}
  {\bibfield  {journal} {\bibinfo  {journal} {Phys. Rev. B}\ }\textbf {\bibinfo
  {volume} {57}},\ \bibinfo {pages} {7846} (\bibinfo {year}
  {1998})}\BibitemShut {NoStop}%
\bibitem [{\citenamefont {Pfeuty}(1970)}]{Pfeuty70:57}%
  \BibitemOpen
  \bibfield  {author} {\bibinfo {author} {\bibfnamefont {P.}~\bibnamefont
  {Pfeuty}},\ }\href@noop {} {\bibfield  {journal} {\bibinfo  {journal} {Annals
  of Physics}\ }\textbf {\bibinfo {volume} {57}},\ \bibinfo {pages} {79}
  (\bibinfo {year} {1970})}\BibitemShut {NoStop}%
\bibitem [{\citenamefont {Schron}\ \emph {et~al.}(2012)\citenamefont {Schron},
  \citenamefont {Rodl},\ and\ \citenamefont {Bechstedt}}]{Schron12:86}%
  \BibitemOpen
  \bibfield  {author} {\bibinfo {author} {\bibfnamefont {A.}~\bibnamefont
  {Schron}}, \bibinfo {author} {\bibfnamefont {C.}~\bibnamefont {Rodl}}, \ and\
  \bibinfo {author} {\bibfnamefont {F.}~\bibnamefont {Bechstedt}},\ }\href@noop
  {} {\bibfield  {journal} {\bibinfo  {journal} {Phys. Rev. B}\ }\textbf
  {\bibinfo {volume} {86}},\ \bibinfo {pages} {115134} (\bibinfo {year}
  {2012})}\BibitemShut {NoStop}%
\bibitem [{\citenamefont {Cabrera}\ \emph {et~al.}(0014)\citenamefont
  {Cabrera}, \citenamefont {Thompson}, \citenamefont {Coldea}, \citenamefont
  {Prabhakaran}, \citenamefont {Bewley}, \citenamefont {Guidi}, \citenamefont
  {Rodriguez-Rivera},\ and\ \citenamefont {Stock}}]{Cabrera14:90}%
  \BibitemOpen
  \bibfield  {author} {\bibinfo {author} {\bibfnamefont {I.}~\bibnamefont
  {Cabrera}}, \bibinfo {author} {\bibfnamefont {J.~D.}\ \bibnamefont
  {Thompson}}, \bibinfo {author} {\bibfnamefont {R.}~\bibnamefont {Coldea}},
  \bibinfo {author} {\bibfnamefont {D.}~\bibnamefont {Prabhakaran}}, \bibinfo
  {author} {\bibfnamefont {R.~I.}\ \bibnamefont {Bewley}}, \bibinfo {author}
  {\bibfnamefont {T.}~\bibnamefont {Guidi}}, \bibinfo {author} {\bibfnamefont
  {J.~A.}\ \bibnamefont {Rodriguez-Rivera}}, \ and\ \bibinfo {author}
  {\bibfnamefont {C.}~\bibnamefont {Stock}},\ }\href@noop {} {\bibfield
  {journal} {\bibinfo  {journal} {Phys. Rev. B}\ }\textbf {\bibinfo {volume}
  {90}},\ \bibinfo {pages} {014418} (\bibinfo {year} {20014})}\BibitemShut
  {NoStop}%
\bibitem [{\citenamefont {Birgeneau}\ \emph {et~al.}(1995)\citenamefont
  {Birgeneau}, \citenamefont {Aharony}, \citenamefont {Belk}, \citenamefont
  {Chou}, \citenamefont {Endoh}, \citenamefont {Greven}, \citenamefont
  {Hosoya}, \citenamefont {Kastner}, \citenamefont {Lee}, \citenamefont {Lee},
  \citenamefont {Shirane}, \citenamefont {Wakimoto}, \citenamefont {Wells},\
  and\ \citenamefont {Yamada}}]{Birgeneau95;56}%
  \BibitemOpen
  \bibfield  {author} {\bibinfo {author} {\bibfnamefont {R.~J.}\ \bibnamefont
  {Birgeneau}}, \bibinfo {author} {\bibfnamefont {A.}~\bibnamefont {Aharony}},
  \bibinfo {author} {\bibfnamefont {N.~R.}\ \bibnamefont {Belk}}, \bibinfo
  {author} {\bibfnamefont {F.~C.}\ \bibnamefont {Chou}}, \bibinfo {author}
  {\bibfnamefont {Y.}~\bibnamefont {Endoh}}, \bibinfo {author} {\bibfnamefont
  {M.}~\bibnamefont {Greven}}, \bibinfo {author} {\bibfnamefont
  {S.}~\bibnamefont {Hosoya}}, \bibinfo {author} {\bibfnamefont {M.~A.}\
  \bibnamefont {Kastner}}, \bibinfo {author} {\bibfnamefont {C.~H.}\
  \bibnamefont {Lee}}, \bibinfo {author} {\bibfnamefont {Y.~S.}\ \bibnamefont
  {Lee}}, \bibinfo {author} {\bibfnamefont {G.}~\bibnamefont {Shirane}},
  \bibinfo {author} {\bibfnamefont {S.}~\bibnamefont {Wakimoto}}, \bibinfo
  {author} {\bibfnamefont {B.~O.}\ \bibnamefont {Wells}}, \ and\ \bibinfo
  {author} {\bibfnamefont {K.}~\bibnamefont {Yamada}},\ }\href@noop {}
  {\bibfield  {journal} {\bibinfo  {journal} {J. Phys. Chem. Solids}\ }\textbf
  {\bibinfo {volume} {56}},\ \bibinfo {pages} {1913} (\bibinfo {year}
  {1995})}\BibitemShut {NoStop}%
\bibitem [{\citenamefont {Greven}\ \emph {et~al.}(1995)\citenamefont {Greven},
  \citenamefont {Birgeneau}, \citenamefont {Endow}, \citenamefont {Kastner},
  \citenamefont {Matsuda},\ and\ \citenamefont {Shirane}}]{Greven95:96}%
  \BibitemOpen
  \bibfield  {author} {\bibinfo {author} {\bibfnamefont {M.}~\bibnamefont
  {Greven}}, \bibinfo {author} {\bibfnamefont {R.~J.}\ \bibnamefont
  {Birgeneau}}, \bibinfo {author} {\bibfnamefont {Y.}~\bibnamefont {Endow}},
  \bibinfo {author} {\bibfnamefont {M.~A.}\ \bibnamefont {Kastner}}, \bibinfo
  {author} {\bibfnamefont {M.}~\bibnamefont {Matsuda}}, \ and\ \bibinfo
  {author} {\bibfnamefont {G.}~\bibnamefont {Shirane}},\ }\href@noop {}
  {\bibfield  {journal} {\bibinfo  {journal} {Z. Phys.}\ }\textbf {\bibinfo
  {volume} {96}},\ \bibinfo {pages} {465} (\bibinfo {year} {1995})}\BibitemShut
  {NoStop}%
\bibitem [{\citenamefont {Makivic}\ and\ \citenamefont
  {Ding}(1991)}]{Makivic91:43}%
  \BibitemOpen
  \bibfield  {author} {\bibinfo {author} {\bibfnamefont {M.~S.}\ \bibnamefont
  {Makivic}}\ and\ \bibinfo {author} {\bibfnamefont {H.~Q.}\ \bibnamefont
  {Ding}},\ }\href@noop {} {\bibfield  {journal} {\bibinfo  {journal} {Phys.
  Rev. B}\ }\textbf {\bibinfo {volume} {43}},\ \bibinfo {pages} {3562}
  (\bibinfo {year} {1991})}\BibitemShut {NoStop}%
\bibitem [{\citenamefont {Hasenfratz}\ and\ \citenamefont
  {Niedermayer}(1991)}]{Hasenfratz83:50}%
  \BibitemOpen
  \bibfield  {author} {\bibinfo {author} {\bibfnamefont {P.}~\bibnamefont
  {Hasenfratz}}\ and\ \bibinfo {author} {\bibfnamefont {F.}~\bibnamefont
  {Niedermayer}},\ }\href@noop {} {\bibfield  {journal} {\bibinfo  {journal}
  {Phys. Lett. B}\ }\textbf {\bibinfo {volume} {268}},\ \bibinfo {pages} {231}
  (\bibinfo {year} {1991})}\BibitemShut {NoStop}%
\bibitem [{\citenamefont {Elliott}\ and\ \citenamefont
  {Thorpe}(1968)}]{Elliott68:39}%
  \BibitemOpen
  \bibfield  {author} {\bibinfo {author} {\bibfnamefont {R.~J.}\ \bibnamefont
  {Elliott}}\ and\ \bibinfo {author} {\bibfnamefont {M.~F.}\ \bibnamefont
  {Thorpe}},\ }\href@noop {} {\bibfield  {journal} {\bibinfo  {journal} {J.
  Appl. Phys.}\ }\textbf {\bibinfo {volume} {39}},\ \bibinfo {pages} {802}
  (\bibinfo {year} {1968})}\BibitemShut {NoStop}%
\bibitem [{\citenamefont {Wilson}\ \emph
  {et~al.}(2007{\natexlab{a}})\citenamefont {Wilson}, \citenamefont
  {Petrenko},\ and\ \citenamefont {Balakrishnan}}]{Wilson07:19}%
  \BibitemOpen
  \bibfield  {author} {\bibinfo {author} {\bibfnamefont {N.~R.}\ \bibnamefont
  {Wilson}}, \bibinfo {author} {\bibfnamefont {O.~A.}\ \bibnamefont
  {Petrenko}}, \ and\ \bibinfo {author} {\bibfnamefont {G.}~\bibnamefont
  {Balakrishnan}},\ }\href@noop {} {\bibfield  {journal} {\bibinfo  {journal}
  {J. Phys. Condens. Matter}\ }\textbf {\bibinfo {volume} {19}},\ \bibinfo
  {pages} {145257} (\bibinfo {year} {2007}{\natexlab{a}})}\BibitemShut
  {NoStop}%
\bibitem [{\citenamefont {Wilson}\ \emph
  {et~al.}(2007{\natexlab{b}})\citenamefont {Wilson}, \citenamefont
  {Petrenko},\ and\ \citenamefont {Chapon}}]{Wilson07:75}%
  \BibitemOpen
  \bibfield  {author} {\bibinfo {author} {\bibfnamefont {N.~R.}\ \bibnamefont
  {Wilson}}, \bibinfo {author} {\bibfnamefont {O.~A.}\ \bibnamefont
  {Petrenko}}, \ and\ \bibinfo {author} {\bibfnamefont {L.~C.}\ \bibnamefont
  {Chapon}},\ }\href@noop {} {\bibfield  {journal} {\bibinfo  {journal} {Phys.
  Rev. B}\ }\textbf {\bibinfo {volume} {75}},\ \bibinfo {pages} {094432}
  (\bibinfo {year} {2007}{\natexlab{b}})}\BibitemShut {NoStop}%
\bibitem [{\citenamefont {Szymczak}\ \emph {et~al.}(2006)\citenamefont
  {Szymczak}, \citenamefont {Baran}, \citenamefont {Diduszko}, \citenamefont
  {Fink-Finowicki}, \citenamefont {Gutowska}, \citenamefont {Szewczyk},\ and\
  \citenamefont {Szymczak}}]{Szymczak06:73}%
  \BibitemOpen
  \bibfield  {author} {\bibinfo {author} {\bibfnamefont {R.}~\bibnamefont
  {Szymczak}}, \bibinfo {author} {\bibfnamefont {M.}~\bibnamefont {Baran}},
  \bibinfo {author} {\bibfnamefont {R.}~\bibnamefont {Diduszko}}, \bibinfo
  {author} {\bibfnamefont {J.}~\bibnamefont {Fink-Finowicki}}, \bibinfo
  {author} {\bibfnamefont {M.}~\bibnamefont {Gutowska}}, \bibinfo {author}
  {\bibfnamefont {A.}~\bibnamefont {Szewczyk}}, \ and\ \bibinfo {author}
  {\bibfnamefont {H.}~\bibnamefont {Szymczak}},\ }\href@noop {} {\bibfield
  {journal} {\bibinfo  {journal} {Phys. Rev. B}\ }\textbf {\bibinfo {volume}
  {73}},\ \bibinfo {pages} {094425} (\bibinfo {year} {2006})}\BibitemShut
  {NoStop}%
\bibitem [{\citenamefont {Petrenko}\ \emph {et~al.}(2010)\citenamefont
  {Petrenko}, \citenamefont {Wilson}, \citenamefont {Balakrishnan},
  \citenamefont {Paul},\ and\ \citenamefont {McIntyre}}]{Petrenko10:82}%
  \BibitemOpen
  \bibfield  {author} {\bibinfo {author} {\bibfnamefont {O.~A.}\ \bibnamefont
  {Petrenko}}, \bibinfo {author} {\bibfnamefont {N.~R.}\ \bibnamefont
  {Wilson}}, \bibinfo {author} {\bibfnamefont {G.}~\bibnamefont
  {Balakrishnan}}, \bibinfo {author} {\bibfnamefont {D.~M.}\ \bibnamefont
  {Paul}}, \ and\ \bibinfo {author} {\bibfnamefont {G.~J.}\ \bibnamefont
  {McIntyre}},\ }\href@noop {} {\bibfield  {journal} {\bibinfo  {journal}
  {Phys. Rev. B}\ }\textbf {\bibinfo {volume} {82}},\ \bibinfo {pages} {104409}
  (\bibinfo {year} {2010})}\BibitemShut {NoStop}%
\bibitem [{\citenamefont {Fritsch}\ \emph {et~al.}(2012)\citenamefont
  {Fritsch}, \citenamefont {Yamani}, \citenamefont {Chang}, \citenamefont
  {Qiu}, \citenamefont {Copley}, \citenamefont {Ramazanoglu}, \citenamefont
  {Dabkowska},\ and\ \citenamefont {Gaulin}}]{Fritsch12:86}%
  \BibitemOpen
  \bibfield  {author} {\bibinfo {author} {\bibfnamefont {K.}~\bibnamefont
  {Fritsch}}, \bibinfo {author} {\bibfnamefont {Z.}~\bibnamefont {Yamani}},
  \bibinfo {author} {\bibfnamefont {S.}~\bibnamefont {Chang}}, \bibinfo
  {author} {\bibfnamefont {Y.}~\bibnamefont {Qiu}}, \bibinfo {author}
  {\bibfnamefont {J.~R.~D.}\ \bibnamefont {Copley}}, \bibinfo {author}
  {\bibfnamefont {M.}~\bibnamefont {Ramazanoglu}}, \bibinfo {author}
  {\bibfnamefont {H.~A.}\ \bibnamefont {Dabkowska}}, \ and\ \bibinfo {author}
  {\bibfnamefont {B.~D.}\ \bibnamefont {Gaulin}},\ }\href@noop {} {\bibfield
  {journal} {\bibinfo  {journal} {Phys. Rev. B}\ }\textbf {\bibinfo {volume}
  {86}},\ \bibinfo {pages} {174421} (\bibinfo {year} {2012})}\BibitemShut
  {NoStop}%
\bibitem [{\citenamefont {Helton}\ \emph {et~al.}(2012)\citenamefont {Helton},
  \citenamefont {Chen}, \citenamefont {Bychokov}, \citenamefont {Barilo},
  \citenamefont {Rogado}, \citenamefont {Cava},\ and\ \citenamefont
  {Lynn}}]{Helton12:24}%
  \BibitemOpen
  \bibfield  {author} {\bibinfo {author} {\bibfnamefont {J.~S.}\ \bibnamefont
  {Helton}}, \bibinfo {author} {\bibfnamefont {Y.}~\bibnamefont {Chen}},
  \bibinfo {author} {\bibfnamefont {G.~L.}\ \bibnamefont {Bychokov}}, \bibinfo
  {author} {\bibfnamefont {S.~N.}\ \bibnamefont {Barilo}}, \bibinfo {author}
  {\bibfnamefont {N.}~\bibnamefont {Rogado}}, \bibinfo {author} {\bibfnamefont
  {R.~J.}\ \bibnamefont {Cava}}, \ and\ \bibinfo {author} {\bibfnamefont
  {J.~W.}\ \bibnamefont {Lynn}},\ }\href@noop {} {\bibfield  {journal}
  {\bibinfo  {journal} {J. Phys.: Condens. Matter}\ }\textbf {\bibinfo {volume}
  {24}},\ \bibinfo {pages} {016003} (\bibinfo {year} {2012})}\BibitemShut
  {NoStop}%
\bibitem [{\citenamefont {Ramazanoglu}\ \emph {et~al.}(2009)\citenamefont
  {Ramazanoglu}, \citenamefont {Adams}, \citenamefont {Clancy}, \citenamefont
  {Berlinsky}, \citenamefont {Yamani}, \citenamefont {Szymczak}, \citenamefont
  {Szymczak}, \citenamefont {Fink-Finowicki},\ and\ \citenamefont
  {Gaulin}}]{Rama09:79}%
  \BibitemOpen
  \bibfield  {author} {\bibinfo {author} {\bibfnamefont {M.}~\bibnamefont
  {Ramazanoglu}}, \bibinfo {author} {\bibfnamefont {C.~P.}\ \bibnamefont
  {Adams}}, \bibinfo {author} {\bibfnamefont {J.~P.}\ \bibnamefont {Clancy}},
  \bibinfo {author} {\bibfnamefont {A.~J.}\ \bibnamefont {Berlinsky}}, \bibinfo
  {author} {\bibfnamefont {Z.}~\bibnamefont {Yamani}}, \bibinfo {author}
  {\bibfnamefont {R.}~\bibnamefont {Szymczak}}, \bibinfo {author}
  {\bibfnamefont {H.}~\bibnamefont {Szymczak}}, \bibinfo {author}
  {\bibfnamefont {J.}~\bibnamefont {Fink-Finowicki}}, \ and\ \bibinfo {author}
  {\bibfnamefont {B.~D.}\ \bibnamefont {Gaulin}},\ }\href@noop {} {\bibfield
  {journal} {\bibinfo  {journal} {Phys. Rev. B}\ }\textbf {\bibinfo {volume}
  {79}},\ \bibinfo {pages} {024417} (\bibinfo {year} {2009})}\BibitemShut
  {NoStop}%
\bibitem [{sup()}]{supp}%
  \BibitemOpen
  \href@noop {} {\bibinfo  {journal} {All data presented in this paper were collected at the ISIS facility (Didcot, UK).  Following UK Research Council guidance and open access policies, the data can either be accessed at source (from ISIS at www.isis.stfc.ac.uk) or through the University of Edinburgh's online digital repository (datashare.is.ed.ac.uk) after publication. }\ }\BibitemShut {NoStop}%
\end{thebibliography}

%

\end{document}